\documentclass[twocolumn]{aastex631}

\newcommand{\wtps}{WTP\,19aalnxx}
\newcommand{\Msun}{M$_\odot$}
\usepackage{comment}
\usepackage{amsmath}

\begin{document}

\title{\wtps: Discovery of a bright mid-infrared transient in the emerging class of low luminosity supernovae revealed by delayed circumstellar interaction}

\author{Charlotte Myers}
\affiliation{MIT-Kavli Institute for Astrophysics and Space Research, 77 Massachusetts Ave., Cambridge, MA 02139, USA}

\author{Kishalay De}
\altaffiliation{NASA Einstein Fellow}
\affiliation{MIT-Kavli Institute for Astrophysics and Space Research, 77 Massachusetts Ave., Cambridge, MA 02139, USA}

\author{Lin Yan}
\affiliation{Cahill Center for Astrophysics, California Institute of Technology, Pasadena, CA 91125, USA}
\affiliation{Caltech Optical Observatories, California Institute of Technology, Pasadena, CA 91125, USA}

\author{Jacob E. Jencson}
\affiliation{Department of Physics and Astronomy, Johns Hopkins University, 3400 North Charles Street, Baltimore, MD 21218, USA}

\author{Nicholas Earley}
\affiliation{Cahill Center for Astrophysics, California Institute of Technology, Pasadena, CA 91125, USA}

\author{Christoffer Fremling}
\affiliation{Cahill Center for Astrophysics, California Institute of Technology, Pasadena, CA 91125, USA}
\affiliation{Caltech Optical Observatories, California Institute of Technology, Pasadena, CA 91125, USA}

\author{Daichi Hiramatsu}
\affiliation{Center for Astrophysics \textbar{} Harvard \& Smithsonian, 60 Garden Street, Cambridge, MA 02138-1516, USA}
\affiliation{The NSF AI Institute for Artificial Intelligence and Fundamental Interactions, USA}

\author{Mansi M. Kasliwal}
\affiliation{Cahill Center for Astrophysics, California Institute of Technology, Pasadena, CA 91125, USA}

\author{Ryan M. Lau}
\affiliation{NSF’s NOIRLab, 950 North Cherry Avenue, Tucson, AZ 85719, USA}

\author{Morgan MacLeod}
\affiliation{Center for Astrophysics \textbar{} Harvard \& Smithsonian, 60 Garden Street, Cambridge, MA 02138-1516, USA}

\author{Megan Masterson}
\affiliation{MIT-Kavli Institute for Astrophysics and Space Research, 77 Massachusetts Ave., Cambridge, MA 02139, USA}

\author{Christos Panagiotou}
\affiliation{MIT-Kavli Institute for Astrophysics and Space Research, 77 Massachusetts Ave., Cambridge, MA 02139, USA}

\author{Robert Simcoe}
\affiliation{MIT-Kavli Institute for Astrophysics and Space Research, 77 Massachusetts Ave., Cambridge, MA 02139, USA}

\author{Samaporn Tinyanont}
\affiliation{National Astronomical Research Institute of Thailand, 260 Moo 4, Donkaew, Maerim, Chiang Mai, 50180, Thailand}

\begin{abstract}
While core-collapse supernovae (SNe) often show early and consistent signs of circumstellar (CSM) interaction, some exhibit delayed signatures due to interaction with distant material around the progenitor star. Here we present the discovery in NEOWISE data of \wtps, a luminous mid-infrared (IR) transient in the outskirts of the galaxy KUG 0022-007 at $\approx 190$\,Mpc. First detected in 2018, \wtps\ reaches a peak absolute (Vega) magnitude of $\approx-22$ at $4.6\,\mu$m in $\approx3$\,yr, comparable to the most luminous interacting SNe. Archival data reveal a $\gtrsim 5\times$ fainter optical counterpart detected since 2015, while follow-up near-IR observations in 2022 reveal an extremely red ($Ks-W2 \approx 3.7$\,mag) active transient. Deep optical spectroscopy confirm strong CSM interaction signatures via intermediate-width Balmer emission lines and coronal metal lines. Modeling the broadband spectral energy distribution, we estimate the presence of $\gtrsim 10^{-2}$\,\Msun\ of warm dust, likely formed in the shock interaction region. Together with the lack of nebular Fe emission, we suggest that \wtps\ is a missed, low (optical) luminosity SN in an emerging family of core-collapse SNe distinguished by their CSM-interaction-powered mid-IR emission that outshines the optical bands. Investigating the Zwicky Transient Facility sample of SNe in NEOWISE data, we find $17$ core-collapse SNe ($\gtrsim 3$\% in a volume-limited sample) without early signs of CSM interaction that exhibit delayed IR brightening, suggestive of dense CSM shells at $\lesssim 10^{17}$\,cm. We suggest that synoptic IR surveys offer a new route to revealing late-time CSM interaction and the prevalence of intense terminal mass loss in massive stars.
\end{abstract}


\section{Introduction} \label{sec:intro}

Massive stars are known to undergo intense terminal mass loss episodes which remain poorly understood \citep{Groh2013, Smith2014}. Not only does this mass loss affect the observational appearance of the subsequent core-collapse supernova (CCSN; e.g. as a hydrogen-free or hydrogen-rich event), the amount, energy and composition of this mass has broader implications for the influence of massive stars on the surrounding interstellar medium \citep{Puls2008}. Elevated mass loss in the years to centuries prior to a massive stellar death can most readily be inferred via the observational effects produced by the interaction of the SN ejecta with the surrounding circumstellar material (CSM; \citealt{Smith2017}). The canonical Type IIn supernovae \citep{Gal-Yam2017} present the most dramatic examples, where narrow and intermediate-width emission lines from the shock interaction region are easily recognized, and the interaction energy powers most of the SN luminosity \citep{Chevalier2017}. However, large-scale optical surveys are now providing increasing evidence that short-lived episodes of dense CSM interaction may be ubiquitous in most CCSNe as revealed by early photometric and spectroscopic follow-up \citep{Bruch2021, Das2023, JG2024}.

While the effects of CSM interaction are sometimes apparent at X-ray and radio wavebands \citep{Chevalier2017}, the mid-infrared (MIR) bands have often provided tantalizing evidence \citep{Szalai2019}. CCSNe frequently show bright MIR emission from warm dust heated by the SN emission \citep{Szalai2013,Tinyanont2016,Fox2011}. Warm dust around CCSNe can be pre-existing in the SN progenitor environment, produced in the dense interaction region between the forward and reverse shocks or even produced in the cooling SN ejecta \citep{Fox2010}. In the presence of dense interaction, the dust can be heated by collisions in the shock or by the radiation produced at the shock. Modeling of the MIR emission as well as its temporal evolution offers powerful clues to understanding the origin of the dust \citep{Fox2011}, while simultaneously providing direct evidence for CSM interaction at late times where optical spectroscopy (to detect interaction narrow lines) may be infeasible \citep{Tinyanont2016, Tinyanont2019, Szalai2021}.

Previous sample studies of MIR supernovae undertaken primarily with the Spitzer space telescope \citep{Fox2011, Tinyanont2016, Szalai2021} show that Type IIn supernovae (see \citealt{Filippenko1997} for a review of SN classification) consistently exhibit luminous mid-IR emission, likely powered by shock radiation reprocessed by dust pre-exisiting in the progentior surroundings in most cases \citep{Fox2011}. Such behavior is less prevalent in the canonical Type IIP SNe; however some events show prolonged MIR emission and even re-brightenings likely due to new dust formation \citep{Fabbri2011, Szalai2011, Meikle2011, Shahbandeh2023} caused by delayed CSM interaction. Similar behavior has also been in some stripped envelope SNe (SESNe) such as SN\,2014C \citep{Tinyanont2019} and SN\,2004dk \citep{Mauerhan2018, Szalai2021} where delayed CSM interaction inferred from optical, radio and X-ray wavelengths coincide with new dust formation. While the delayed MIR emission in these events is comparable or fainter than the primary optical supernova, it is straightforward to envision a population where MIR emission powered by delayed interaction completely outshines its parent low-luminosity SN. However, the lack of synoptic MIR observations outside of targeted follow-up campaigns or serendipitous host galaxy observations severely limit systematic searches in this regime.

Here, we present the discovery and characterization of a luminous MIR transient in NEOWISE data, utilizing the unique wavelength coverage, temporal baseline and cadence of the NEOWISE mission. The transient exhibits the tell-tale signatures of a CSM-interaction powered SN where the original optical emission was missed due to its low luminosity. We present the discovery and observational details in Section \ref{sec:obs}. We present a detailed analysis of the photometric and spectroscopic characteristics in Section \ref{sec:analysis}. In Section \ref{sec:discussion}, we identify additional examples of this class of events in NEOWISE data and conclude with a summary. Throughout this paper, we assume $\Omega_M = 0.3$, $\Omega_\Lambda = 0.7$, and $H_0 = 70$\,km\,s$^{-1}$\,Mpc$^{-1}$. Unless otherwise noted, quoted uncertainties represent 90\% confidence intervals.

\section{Observations}
\label{sec:obs}
\subsection{Discovery in NEOWISE data}

\begin{figure*}
\begin{center}
\includegraphics[width = 0.9\textwidth]{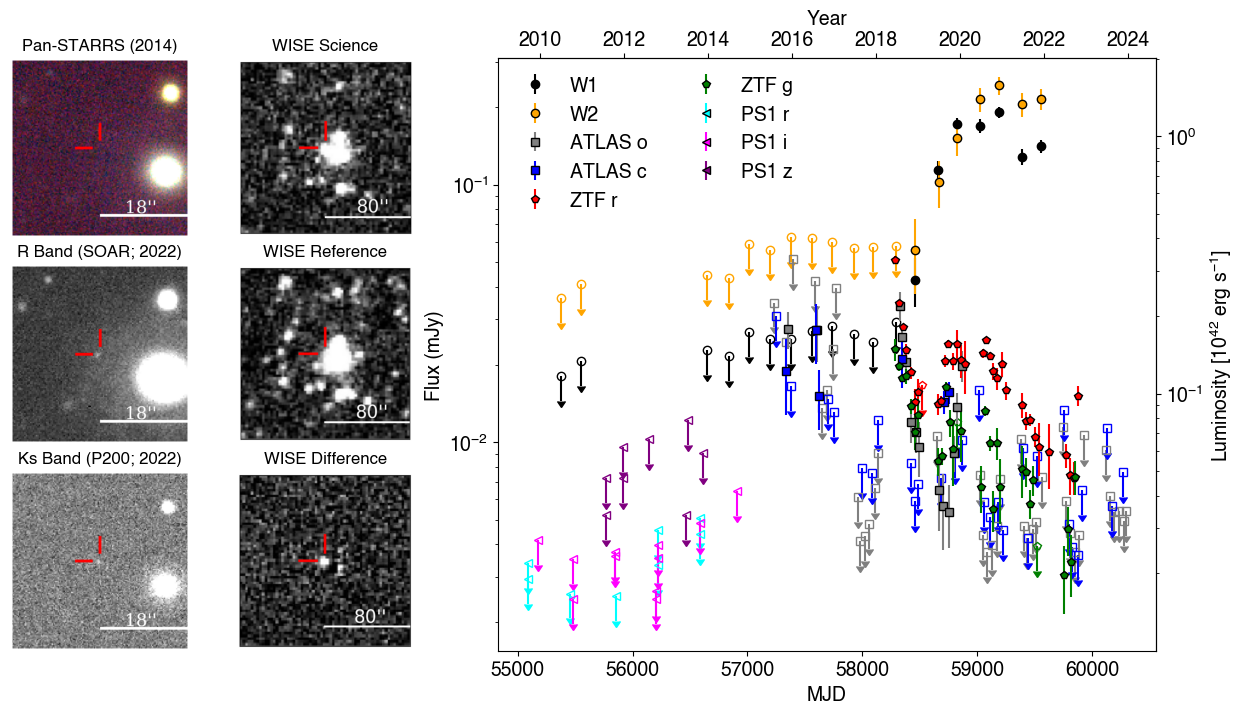}
\caption{The discovery location and multi-wavelength light curve of \wtps. {\it Left panels}: Multi-color optical ($gri$) pre-explosion image from the Pan-STARRS1 survey (top), post-explosion late-time $r$-band imaging using SOAR (middle) and post-explosion late-time $K$-band imaging using the P200 (bottom). {\it Middle panels}: Science, reference and difference imaging triplet of the location of \wtps\ showing a detection of a MIR transient. {\it Right panel}: Multi-wavelength light curve of \wtps\ from NEOWISE, PS1, ATLAS and ZTF surveys. Solid symbols denote detections in the respective band while hollow symbols with downward arrows indicate upper limits.}
\label{fig:lc}
\end{center}
\end{figure*}

The Wide-field Infrared Survey Explorer (WISE) satellite \citep{Wright2010}, re-initiated as the NEOWISE mission \citep{Mainzer2014}, has been carrying out an all-sky MIR survey in the $W1$ ($3.4$\,$\mu$m) and $W2$ ($4.6$\,$\mu$m) bands since 2014. In its ongoing survey, NEOWISE revisits each part of the sky once every $\approx 0.5$\,yr. We have carried out a systematic search for transients in time-resolved coadded images created as part of the unWISE project \citep{Lang2014,Meisner2018}, the details of which will be presented in De et al. (in prep). In brief, we used a customized code \citep{De2019} based on the ZOGY algorithm \citep{Zackay2016} to perform image subtraction on the NEOWISE images using the co-added images of the WISE mission (obtained in 2010-2011) as reference images\footnote{For all transients identified in the WISE Transient Pipeline (WTP, De in prep.), we adopt the naming scheme WTP\,XXYYYYYY, where XX indicates the year of first detection and YYYYYY is a six letter alphabetical code.}. Our pipeline produces a database of all transients down to a statistical significance of $\approx 10\sigma$. Follow-up for the sources was coordinated using the \texttt{fritz} astronomical data platform \citep{vanderWalt2019}.

We identified the transient source \wtps\ at J2000 coordinates $\alpha=$00:24:41.60, $\delta=-$00:30:14.99 in a crossmatch between the sample of NEOWISE transients and galaxies with confirmed spectroscopic redshifts in the Census of the Local Universe (CLU; \citealt{Cook2019}) catalog. The source lies in the outskirts (projected offset $\approx 14.2$\,\arcsec) of the spiral galaxy KUG\,0022-007 at a known redshift of $z = 0.043$. The corresponding luminosity distance is $D_L = 190$\,Mpc and the projected physical offset from the center of the galaxy is $\approx 12$\,kpc, within the spiral arms of the host galaxy. Performing forced difference Point Spread Function (PSF) photometry at the source position in NEOWISE images \citep{De2023}, we find the mid-IR brightening to have begun in $\approx 2018$, reaching a peak flux of $\approx 0.25$\,mJy in the $W2$ band in 2021 before beginning to fade. The complete WISE light curve of the source is shown in Figure \ref{fig:lc}. We convert the measured fluxes on the unWISE images to physical flux units using the published WISE zero-point fluxes\footnote{\url{https://wise2.ipac.caltech.edu/docs/release/allsky/expsup/sec4_4h.html}}.

\subsection{Archival light curves and follow-up observations}

\begin{figure*}[ht!]
\includegraphics[width = \textwidth]{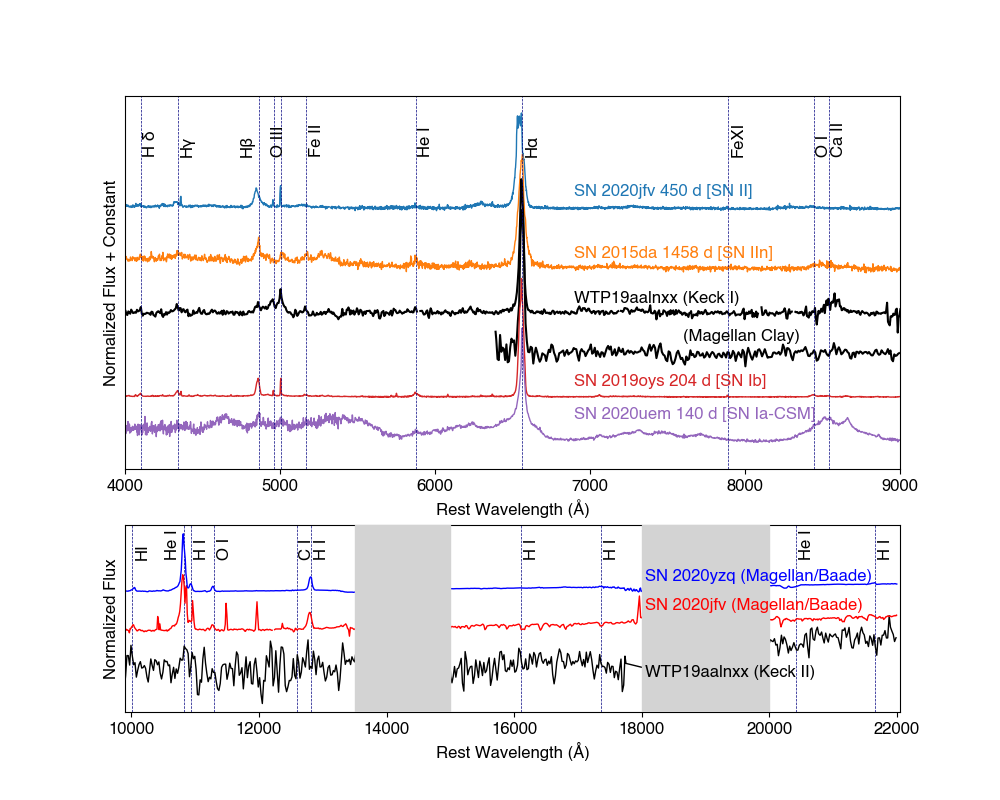}
\caption{Comparison of the late-time spectra of \wtps\ to other known SNe exhibiting late-time CSM interaction. {\it Top panel}: The late-time optical spectrum compared to other known interacting SNe spanning different spectroscopic types at peak light -- the Type IIb (SN 2020jfv), Type IIn (2015da), Type Ib (SN 2019oys) and Type IIP (SN 2020uem) SNe. Bottom: Near-infrared spectrum of \wtps\ compared to our Magellan/FIRE follow-up of two additional optically discovered SNe exhibiting late-time MIR brightening -- the Type IIb (SN\,2020jfv) and Type II (SN\,2020yzq). Regions of low atmospheric transparency have been masked with gray shaded areas.
\label{fig:spec}}
\end{figure*}

We retrieved archival light curves at the position from the Zwicky Transient Facility (ZTF; \citealt{Bellm2019}), Asteroid Terrestrial-impact Last Alert System (ATLAS; \citealt{Tonry2018}) and PanSTARRS surveys \citep{Chambers2016}. The technical details can be found in Appendix \ref{sec:archival}. The combined light curve is shown in Figure \ref{fig:lc}. We obtained follow-up NIR imaging of the source with the Palomar 200-inch telescope and optical/NIR imaging with the 4.1 m Southern Astrophysical Research (SOAR) Telescope. The details of the observations and data reduction can be found in Appendix \ref{sec:followup}. The follow-up images are shown in Figure \ref{fig:lc}. We also obtained follow-up optical spectroscopy of \wtps\ with the Keck-I and Magellan/Clay telescopes, and follow-up NIR spectroscopy of \wtps\ with the Keck-II telescope and two similar bright MIR SNe with the Magellan/Baade telescope, described in Appendix \ref{sec:followup}. The spectra are presented in Figure \ref{fig:spec}.

\section{Analysis}
\label{sec:analysis}
\subsection{Comparison to known supernovae in the mid-IR}
\begin{figure*}
    \centering
    \includegraphics[width=0.99\textwidth]{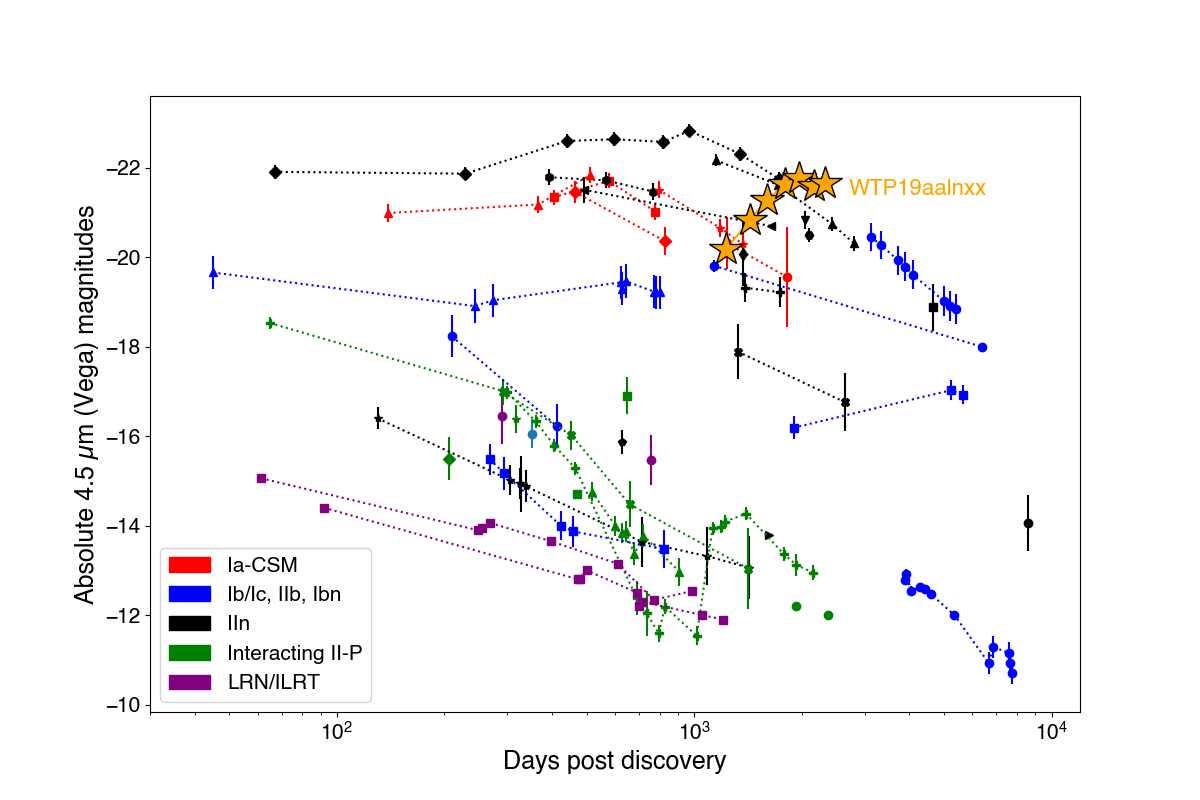}
    \caption{Comparison of the $W2$ light curve of \wtps\ to $4.5\,\mu$m light curves of different types of SNe observed with the Spitzer space telescope \citep{Szalai2021}. SNe of a given type are plotted with the same color (shown in legend) while the evolution of each object is shown with connected symbols.}
    \label{fig:mir_comp}
\end{figure*}

We begin our analysis by comparing the MIR luminosity evolution of \wtps\ with the known population of extragalactic MIR transients. In Figure \ref{fig:mir_comp}, we plot the $W2$ light curve of \wtps\ with the complete sample of $4.5\,\mu$m light curves of supernovae from the \textit{Spitzer} space telescope \citep{Szalai2019, Szalai2021}. The epoch of explosion is however not well constrained; while the WISE brightening begins in 2018, archival optical photometry reveals a faint transient preceding the mid-IR emission. We can constrain the epoch of explosion noting that the PS1 data do not detect any source until late 2014, while the first ATLAS observations in mid-2015 reveal evidence of a faint optical source. We nominally adopt this date (MJD $\approx 57230$) as the explosion epoch, noting that it may be uncertain by $\lesssim 300$\,d due to the gap in coverage between PS1 and ATLAS. The optical source fades below detectability after the initial ATLAS detection within $\approx 6$\,months, but subsequently re-brightens during 2018 as detected in both ZTF and ATLAS data and continues to exhibit a bumpy morphology until late 2022. We highlight a sudden reddening in the $g-r$ color at the onset of the MIR brightening in 2019, likely produced by the strong H$\alpha$ line starting to dominate the source spectrum (Section \ref{sec:spec}).

\wtps\ reaches a peak absolute Vega magnitude of $\lesssim -21.5$ ($W2$), lying amongst the most luminous known mid-IR supernovae. The high luminosity is unlike that seen in SN impostors such as Luminous red novae (LRNe; \citealt{Pastorello2019}) and Intermediate Luminosity Red Transients (ILRTs; \citealt{Cai2021}), as well as Type IIP and Ib/c SNe. The luminosity and slow evolution is only similar to Type IIn and Type Ia-CSM SNe that are powered by circumstellar interaction. Under the assumption that the faint ATLAS emission is coincident with the start of the terminal explosion, the MIR brightening begins $\gtrsim 1000$\,d after the explosion, unlike all the IIn/Ia-CSM SNe; instead the brightening timescale is similar to that seen in the subset of Type Ib/c SNe which exhibit late-time MIR re-brightening and fading, suggested to arise from late-time CSM interaction \citep{Szalai2021, Tinyanont2016}. We conclude that the high $W2$ luminosity and brightening timescale of \wtps\ is indicative of a MIR transient powered by late-time CSM interaction.

\subsection{Spectroscopic signatures}
\label{sec:spec}

Given the MIR evidence for interaction and dust in \wtps\ we compare its optical and NIR spectrum to that of other well-studied SNe exhibiting late-time CSM interaction. In Figure \ref{fig:spec}, we compare the late-time optical spectrum with two well studied interacting SNe from the literature -- the Type IIn SN\,2015da \citep{Tartaglia2020} and the Type\,Ia-CSM SN\,2020uem \citep{Sharma2023}. \wtps\ exhibits all the characteristic optical spectroscopic signatures of an interacting SN --
narrow/intermediate-width lines of the Balmer series (H$\alpha$, H$\beta$, H$\gamma$) and He\,I  (5876 \r{A}) emission, together with coronal metal lines of Fe. We also observe resolved lines of [O\,III] (4959 \r{A}, 5007 \r{A}), commonly seen in other interacting SNe. Fitting the strong H$\alpha$ emission line, we find that it is well described by a sum of two Gaussian profiles, consisting of a resolved, intermediate-width component with a full-width-at-half-maximum of $\approx 1200$\,km\,s$^{-1}$ superimposed with a broad component with a width of $\approx 4400$\,km\,s$^{-1}$. Unlike the optical spectrum, the NIR spectrum of \wtps\ exhibits a red featureless continuum without clear signs of emission lines; we discuss possible interpretations in Section \ref{sec:discussion}.

We note that the broad Fe group emission lines seen in Ia-CSM SNe at nearly all phases \citep{Sharma2023} are not observed in \wtps. Together with the long and bumpy optical light curve (Figure \ref{fig:lc}), the observations argue against a Ia-CSM classification for this object. While the the spectrum has remarkable similarities to that of SN\,2015da, the lower luminosity optical light curve does not support a SN\,IIn classification; most well-studied SN\,IIn peak within $-20.0 \lesssim M \lesssim -18.0$ \citep{Nyholm2020}, while \wtps\ peaks at $M_{r}$ = -16.7 in the optical. Given that the emission is dominated by the CSM interaction at this late stage (and not by the photospheric lines that are typically used to classify SNe), we compare the optical spectrum to other SN types that exhibit delayed CSM interaction. Figure \ref{fig:spec} shows that the spectrum is also broadly similar to other SNe that exhibit delayed CSM interaction despite being classified as a different spectroscopic sub-type near peak optical light -- including the Type\,Ib SN\,2019oys \citep{Sollerman2020} and the Type II SN\,2020jfv \citep{Sollerman2021}. As such, we suggest that \wtps\ is a likely low-luminosity core-collapse SN exhibiting delayed CSM interaction, and consistent with both stripped envelope and hydrogen-rich SNe \citep{Spiro2014}. However, the spectroscopic type near peak light is unconstrained. We can constrain the mass-loss rate of the pre-existing CSM \citep{Smith2017} assuming the interaction energy is primarily re-radiated via the bright H$\alpha$ emission line ($L_{\rm H\alpha} \approx 10^{40}$\,erg\,s$^{-1}$, as calibrated to $r$-band photometry) as
\begin{equation}
\begin{split}
    \dot{M} \approx 2.5 \times 10^{-5} {\rm M_\odot\,yr^{-1}} \left( \frac{L_{\rm H\alpha}}{10^{40}\,{\rm erg\,s}^{-1}} \right) \left(\frac{V_{CSM}}{100\,{\rm km\,s}^{-1}}\right)\\ \left(\frac{V_{CDS}}{5000\,{\rm km\,s}^{-1}}\right)^{-3}
\end{split}
\end{equation}
where $V_{CSM}$ is the assumed wind velocity and $V_{CDS}$ is the velocity of the ejecta interaction region. The derived mass-loss rate at a time $t_{exp}$ after the explosion corresponds to a time
\begin{equation}
    t \approx 270\,{\rm yr} \left( \frac{t_{exp}}{2000\,{\rm d}}\right) \left(\frac{V_{CDS}}{5000\,{\rm km\,s}^{-1}}\right) \left(\frac{V_{CSM}}{100\,{\rm km\,s}^{-1}}\right)^{-1}
\end{equation}
before the death of the star, thereby tracing terminal mass loss in the centuries before the explosion.

\subsection{Dust temperature and mass}

\begin{figure*}[!ht]
    \centering
    \includegraphics[width=0.54\textwidth]{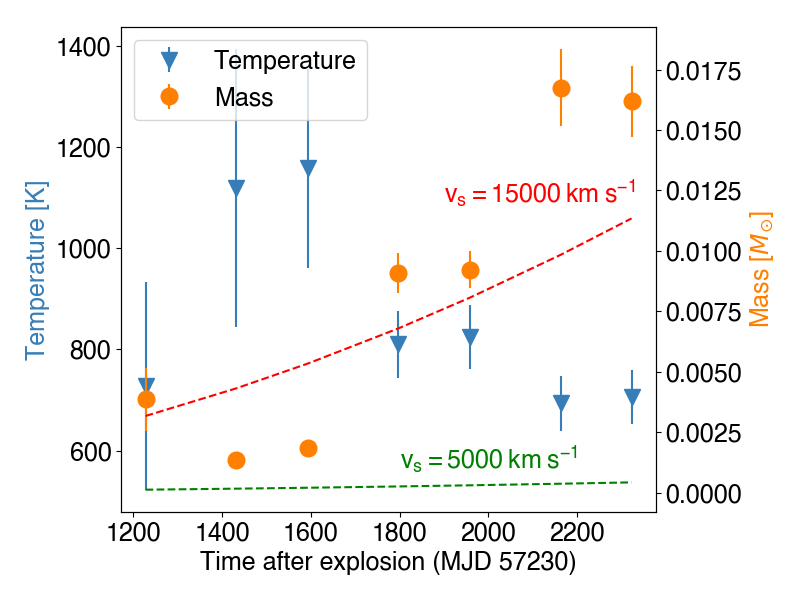}
    \includegraphics[width=0.45\textwidth]{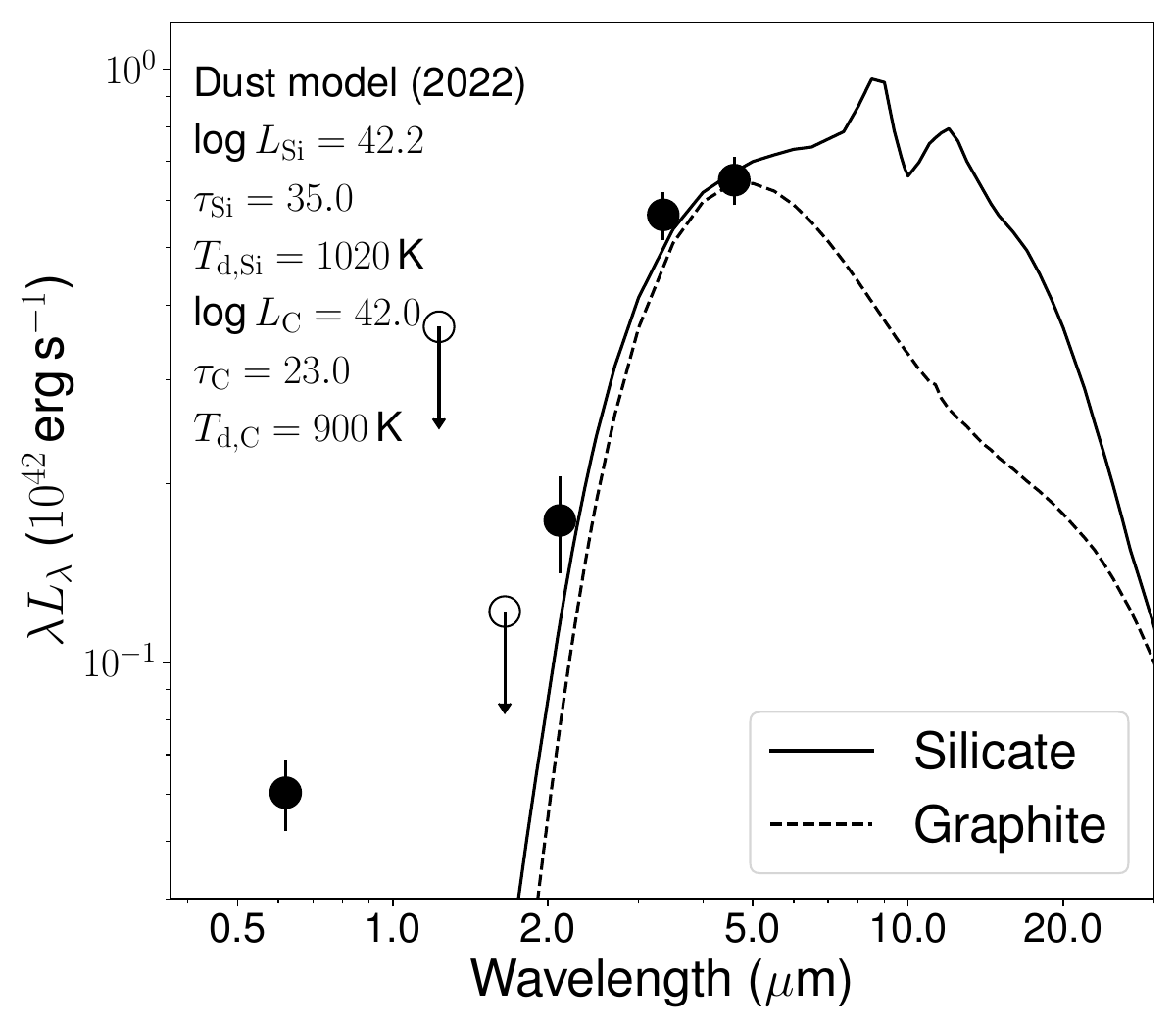}
    \caption{Modeling of the spectral energy distribution of \wtps. {\it Left}: Temporal evolution of the temperature and mass of dust inferred from fitting the NEOWISE MIR photometry (see text). The dashed lines show the expected dust mass in the case of collisional shock heating in the interaction region, for two different ejecta velocities \citep{Fox2010}. {\it Right}: Best-fit models of the SED from $r$-band to $W2$-band in 2022 obtained using \texttt{DUSTY}. We show the best-fit models for both silicate (denoted as Si) and graphite (denoted as C) dust compositions; the best-fit parameters for the two cases are shown in the text.}
    \label{fig:dust_model}
\end{figure*}

The luminous and variable MIR emission of \wtps\ is indicative of a hot dust shell surrounding the likely central SN. We attempt to constrain the temporal evolution of the temperature and mass of the shell by fitting the MIR WISE photometry with a simple blackbody model accounting for the wavelength-dependent emissivity of the dust grains. As in \citet{Fox2010} and \citet{Fox2011}, we nominally use the emissivity model for silicate dust with $a = 0.1\,\mu$m grains from \citet{Draine2001} and fit for the temperature and mass using the modified Planck function. The mass of the dust is derived from the best-fit model as
\[ M = \frac{(4/3)a \rho_b  D^2 F_\nu }{Q_S(\lambda, a) B_\nu (T_d)}\]
where \textit{a} is the assumed grain size of the dust (0.1\,$\mu$m), $\rho_b$ is the bulk density ($2.2$\,g\,cm$^{-3}$), \textit{D} is the distance to the source, $F_\nu$ is the observed flux, $Q_S$ is the grain emissivity factor based on the model for silicate dust grains, and $B_\nu$ is the Planck function at the estimated temperature. Figure \ref{fig:dust_model} shows the temporal evolution the dust temperature and mass. The model assumes that the dust emission is optically thin and therefore provides a lower limit to the true dust mass for the given composition and grain size.

The temperature increases to a maximum of $\approx$1200 K within $\approx 1$\,yr of the appearance of the MIR transient, subsequently cooling to $\lesssim 700$\,K over the next $\approx 1000$\,d. At the same time, the inferred dust mass grows over time from $\approx 10^{-3}$\,\Msun to $\gtrsim 10^{-2}$\,\Msun, potentially suggestive of new dust formation in the ejecta. We derive similar estimates for the evolution of the dust mass if we instead assume $0.1\,\mu$m graphitic dust, albeit at lower temperatures (cooling from $\approx 900$\,K to $\approx 600$\,K).  We draw additional constraints on the nature of the dust emission noting that the inferred blackbody radius for a pure blackbody model provides a lower limit on the true radius of the shell \citep{Szalai2011, Fox2010}. The inferred radius increases from $\approx 3 \times 10^{16}$\,cm to $\approx 8 \times 10^{16}$\,cm over the same time period, suggesting an expansion velocity of $\approx 6000$\,km\,s$^{-1}$ similar to that measured for the broad H$\alpha$ component in the optical spectrum. If the explosion powering the dust emission began $\approx 1000$\,d before the MIR brightening (Figure \ref{fig:mir_comp}), the inferred spectroscopic velocity would suggest that the ejecta radius would also evolve from $\approx 4 \times 10^{16}$\,cm to $\approx 8\times 10^{16}$\,cm over the duration where the source is detected in WISE.

We revisit the assumption that the dust emission is optically thin by modeling the broadband spectral energy distribution (SED) with optical and near-infrared photometry. We use our only optical ($r$-band) and NIR ($H$ and $K$ band) constraints of the source from 2022 and fit it together with the MIR photometry using \texttt{DUSTY} \citep{Ivezic1997, Ivezic1999}. We assume a spherically symmetric distribution of the dust with a $\propto r^{-2}$ density profile around the central explosion and a shell outer to inner radius of $Y = 2$. We consider both Silicate and Graphite-rich compositions for the dust with a MRN grain size distribution (\citealt{Mathis1977}; $\propto a^{-3.5}$) with a minimum and maximum grain size of $a_{\rm min} = 0.005$\,$\mu$m and $a_{\rm max} = 0.25$\,$\mu$m. We fit the photometry using a Markov Chain Monte Carlo (MCMC) wrapper around the \texttt{DUSTY} code \citep{De2022} using the Python \texttt{emcee} library \citep{Foreman-Mackey2013}. The resulting free parameters of the model are the dust optical depth at $0.55\,\mu$m ($\tau_V$), the inner stellar temperature ($T_*$), the dust temperature at the inner edge of the shell ($T_d$) and the total flux ($F$). We assume flat priors on all the fit parameters and ensure convergence of the posterior sampling chains.

The resulting fits are shown in Figure \ref{fig:dust_model}. We note that the 2022 $r$-band photometry is significantly in excess of the SED models given the spectral slope from the NEOWISE to NIR bands. As discussed in Section \ref{sec:spec}, the $r$-band photometry is clearly contaminated with a strong H$\alpha$ emission line that is not part of the continuum, and therefore we exclude the $r$-band data in the final fit. The best-fit models indicate a high optical depth of $\tau \approx 25-35$ together with a dust temperature of $\approx 800 - 1100$\,K, similar to that inferred from the modified blackbody fitting. Both the high optical depths and the lack of longer wavelength coverage (which could allow large masses of cold dust to be missed) suggest that the derived dust masses are lower limits to the true mass. The \texttt{DUSTY} models also constrain the total luminosity of the dust shell to be $\log L \approx 42.0$, consistent with a luminous MIR transient. While both the silicate and graphite dust models can fit with the near to mid-IR photometry, they exhibit dramatic differences at $\lambda > 5\,\mu$m. These differences can be used to accurately constrain the dust composition with longer wavelength spectroscopy or photometry.

\section{Discussion}
\label{sec:discussion}

\subsection{A luminous MIR transient likely powered by dust formation in a shock}

In Section \ref{sec:analysis}, we show that \wtps\ lands among the most luminous MIR transients known in the literature, comparable only to events powered by CSM interaction. The delayed brightening of the MIR emission compared to the optical SN (at $\gtrsim 1000$\,d) is similar to that suggested for core-collapse SNe exhibiting delayed interaction with distant CSM shells. Comparison of the late-time optical spectra confirm striking similarities with previously known core-collapse SNe exhibiting late-time CSM interaction. The very high inferred mass-loss rate from the optical spectra ($\dot{M} \approx 2.5 \times 10^{-5}$\,M$_\odot$\,yr$^{-1}$) is similar to that observed in interacting SNe and the winds of Luminous Blue Variables \citep{Smith2017}.

We now attempt to constrain the nature of the MIR emission. The delayed brightening rules out an IR echo scenario where the MIR emission can be sustained due to light travel time effects of a pre-existing dust shell heated by the original CSM around the progenitor. The inferred minimum radius of the dust shell from the MIR photometry is consistent the expected location of the expanding ejecta at the time of the MIR brightening. While it is possible that the MIR brightening is caused by new dust formation within the cooling ejecta, we consider it unlikely due to the temporal coincidence of the MIR brightening with the onset of undulations in the optical light curve (Figure \ref{fig:lc}), indicative of CSM interaction triggering the dust emission.

If the dust is pre-existing, we use the relations in \citet{Fox2010} for the dust mass expected to be collisionally heated in the shock, and find that the measured masses exceed the model by $\gtrsim 10\times$ for the likely velocity ($\approx 5000$\,km\,s$^{-1}$; Figure \ref{fig:dust_model}), and even in the case of a velocity of $15000$\,km\,s$^{-1}$. Radiative heating of a pre-existing dust shell due to the optical brightening of the SN (Figure \ref{fig:lc}; likely due to the onset of CSM interaction) cannot be ruled out; however it is unclear why the prior optical peak (only $\approx 2\times$ fainter) did not produce detectable MIR emission. Another possibility is that pre-existing dust is heated due to the onset of CSM interaction at late-times, in which case the dust should lie outside the region where the ejecta is interacting with CSM. However, given the striking similarity of the estimated radius of the ejecta (measured from spectroscopy and light curve) and the dust formation radius (estimated from the MIR SED) around the time of the MIR brightening, it appears likely that new dust formation took place in the dense interaction region \citep{Smith2017} consistent with the increasing dust mass (Figure \ref{fig:dust_model}).

\subsection{An emerging population of MIR SNe outshining the optical bands}

\begin{figure*}
    \centering
    \includegraphics[width=0.9\textwidth]{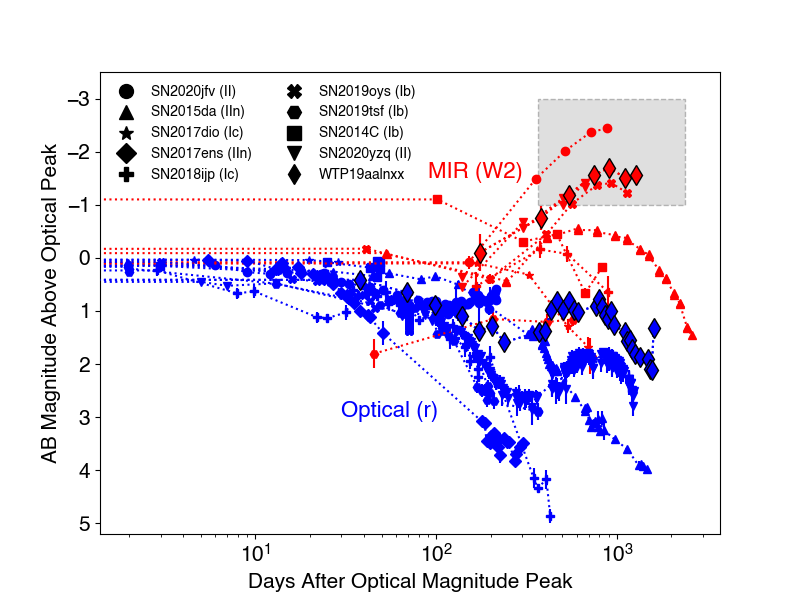}
    \caption{Comparison of the optical and MIR light curves for a sample of SNe, selected to include known SNe exhibiting signs of early or late-time CSM interaction. The comparison sample includes the Type II SN\,2020jfv \citep{Sollerman2021}, the Type Ib SN\,2019oys \citep{Sollerman2020}, the Type IIn SN\,2015da \citep{Tartaglia2020}, the Type Ib SN\,2019tsf \citep{Sollerman2020}, the Type Ic SN\,2017dio \citep{Kuncarayakti2018}, the Type IIn SN\,2014C \citep{Tinyanont2016}, the Type Ic-BL SN\,2017ens \citep{Chen2018}, the Type Ic-BL SN\,2018ijp \citep{Tartaglia2020} and the Type II SN\,2020yzq \citep{Zhang2020}. The optical light curves were obtained from the respective publications, and the MIR $W2$ light curves were obtained from our forced photometry pipeline. While the majority of events exhibit sustained bright MIR emission comparable in flux to the optical peak, a small sample of sources exhibit dramatic MIR brightenings at $\gtrsim 1$\,year after and $> 1$\,mag brighter than the optical peak (including \wtps; shown in the gray shaded region).}
    \label{fig:opt_mir_space}
\end{figure*}

One of the most striking features of \wtps\ is its discovery as a luminous MIR transient that was completely missed in optical searches. Despite its proximity in the local universe ($< 200$\,Mpc), it was not spectroscopically classified or reported due to its faint peak optical magnitude of $r\approx 19.6$\,mag. While ongoing systematic SN classification experiments (e.g. \citealt{Fremling2020, Neumann2023}) are not sensitive to low luminosity core-collapse SNe at this distance (\wtps\ exhibits a peak absolute magnitude of $M_r \approx -16.7$), \wtps\ is distinguished by its exceptional MIR emission that outshines the optical SN by $\gtrsim 1.5$\,mag. Noting that the likely interpretation of the MIR emission is due to late-time CSM interaction, we searched for known interacting SNe published in the literature to confirm similar behavior. We show a collage of the optical and MIR light curves of some published events exhibiting CSM interaction in Figure \ref{fig:opt_mir_space}.

While most interaction-powered events exhibit MIR emission that is sustained in time and similar in brightness to the optical peak, we highlight a sub-sample of events (SN\,2020jfv, SN\,2019oys, SN\,2020yzq and \wtps) that exhibit dramatic MIR brightening at $\gtrsim 1$\,yr after the optical peak that outshines the optical peak by $\gtrsim 1$\,mag ($\gtrsim 2.5\times$ in flux). All these events exhibit faint late-time rebrightening episodes in their optical light curves. To compare the IR spectra of these events to \wtps\, we show new NIR spectra of SN\,2020jfv and SN\,2020yzq in Figure \ref{fig:spec}. Both objects show clear signs of CSM interaction in their spectra via intermediate-width emission lines of H, He and O; however such features are not seen in the NIR spectra of \wtps, and we cannot rule out if the NIR spectroscopic interaction signatures in \wtps\ may have appeared at later epochs (the optical spectrum was obtained nearly a year after the NIR). A similar lack of optical spectroscopic signatures was reported for SN\,2019tsf \citep{Sollerman2020}, which like \wtps, also exhibits a late-time MIR brightening (Figure \ref{fig:opt_mir_space}).

\subsection{The prevalence of core-collapse SNe exhibiting late MIR brightening}

\wtps\ demonstrates that MIR studies offer a powerful probe to reveal late-time CSM interaction in core-collapse SNe even in the absence of expensive late-time optical spectroscopy (usually carried out on 6-10\,m class telescopes). While previous studies of late-time MIR emission in core-collapse SNe have been primarily limited to targeted or serendipitous observations of nearby galaxies with the Spitzer space telescope \citep{Szalai2021}, the NEOWISE dataset is uniquely suited to discover events exhibiting late-time CSM interaction even in the absence of spectroscopic signatures at peak light and late times. To quantify the prevalence of MIR brightening possibly caused by CSM interaction, we use the systematic sample of spectroscopically classified core-collapse SNe reported by the Zwicky Transient Facility Bright Transient Survey (BTS; \citealt{Fremling2020, Perley2020}). The experiment is complete for SNe brighter than $18.5$\,mag at peak light, and we use it to construct a controlled sample of optically discovered events between 2018-2020 (735 total SNe; ensuring $>2$ years of NEOWISE data after the optical SN).

Performing forced photometry in the NEOWISE difference images (as described in Section \ref{sec:obs}), we visually search for events that i) exhibit late MIR brightening in either $W1$ or $W2$ bands after the fading of the optical SN and ii) are not spectroscopically classified as interacting SNe on the Transient Name Server (typically based on peak light spectroscopy) to demonstrate the utility of MIR observations in identifying late-time CSM interaction. We identify a sample of 17 core-collapse SNe exhibiting such signatures and show their optical and MIR light curves in Appendix \ref{sec:mir_bright}. While the majority of events fade below detectability in the optical bands by the time of the MIR brightening, some events indeed show late time optical plateaus and rebrightening at flux levels $\gtrsim 1-2$\,mags fainter than the optical peak, consistent with their interpretation as interacting SNe. Constructing a volume-limited sub-sample of events with $z < 0.036$ (corresponding to a complete sample of SNe brighter than $M = -17.5$\,mag at the depth of BTS), we find a lower limit\footnote{The derived numbers represent a lower limit due to the six-month NEOWISE cadence, where faint late-time re-brightening episodes near the detection threshold may be missed in the observing gap.} of $\gtrsim 3$\% of core-collapse SNe (12 out of 495) without early spectroscopic interaction signatures exhibit late MIR brightening with $m_{\rm AB} \lesssim 19.5$\,mag (the NEOWISE sensitivity) at $\lesssim 1000$\,d (corresponding to CSM located at $\lesssim 10^{17}$\,cm for a shock velocity of $10^4$\,km\,s$^{-1}$) from optical peak.

\section{Summary}
\label{sec:summary}

We present the discovery and characterization of \wtps\, a luminous MIR transient in the outskirts of a spiral galaxy at $\approx 190$\,Mpc. Based on i) the timescale and luminosity of its MIR light curve, ii) faint optical light curve exhibiting multiple rebrightening episodes, iii) spectroscopic signatures of CSM interaction without nebular Fe emission and iv) temporal evolution of the dust temperature and mass derived from the MIR light curve, we suggest that \wtps\ is a low luminosity core-collapse SN exhibiting dramatic MIR brightening powered by delayed CSM interaction. By inspection of known interacting SNe from the literature, we highlight an emerging population of core-collapse SNe exhibiting delayed MIR brightening episodes that completely outshine the parent optical SN. Not only do these episodes offer a completely new route to both discovering massive stellar death in the local universe, they offer direct evidence for likely detached CSM shells around the progenitors of core-collapse SNe possibly arising from eruptive mass loss in the decades to centuries prior to explosion.

Investigating the spectroscopically complete sample of nearby core-collapse SNe from the Zwicky Transient Facility, we identify a sample of 17 events that exhibit late-time MIR brightening episodes that do not otherwise exhibit spectroscopic interaction signatures at peak light. Our findings suggest that $\gtrsim 3$\% of core-collapse SNe exhibit delayed CSM interaction signatures; while a detailed analysis confirming whether these brightening episodes are consistent with CSM interaction is beyond the scope of this paper, such searches can motivate targeted optical spectroscopic follow-up to confirm as well as measure the composition and dynamics of the CSM shells. Longer wavelength follow-up with the JWST offers the only opportunity to accurately measure the dust chemistry and distribution (Figure \ref{fig:dust_model}). Finally, upcoming synoptic IR surveys from the ground like WINTER \citep{Lourie2020} and PRIME \citep{Kondo2023} as well as in space, like the {\it NEO-Surveyor} mission \citep{Mainzer2023} and {\it Roman} high latitude time domain survey (e.g. \citealt{Wang2023}), will open new windows into revealing this population owing to both their increased sensitivity as well as the larger detectable volume for interacting IR SNe that outshine their optical counterparts.

\section*{Acknowledgments}.

We thank Y. Sharma for providing the optical spectra of SNe\,Ia-CSM events. K. D. was supported by NASA through the NASA Hubble Fellowship grant \#HST-HF2-51477.001 awarded by the Space Telescope Science Institute, which is operated by the Association of Universities for Research in Astronomy, Inc., for NASA, under contract NAS5-26555. This work was supported by a NASA Keck PI Data Award, administered by the NASA Exoplanet Science Institute. Data presented herein were obtained at the W. M. Keck Observatory from telescope time allocated to the National Aeronautics and Space Administration through the agency's scientific partnership with the California Institute of Technology and the University of California. The Observatory was made possible by the generous financial support of the W. M. Keck Foundation. The authors wish to recognize and acknowledge the very significant cultural role and reverence that the summit of Maunakea has always had within the indigenous Hawaiian community. We are most fortunate to have the opportunity to conduct observations from this mountain. This research has made use of the Keck Observatory Archive (KOA), which is operated by the W. M. Keck Observatory and the NASA Exoplanet Science Institute (NExScI), under contract with the National Aeronautics and Space Administration. This paper includes data gathered with the 6.5 meter Magellan Telescopes located at Las Campanas Observatory, Chile. Based in part on observations obtained at the Southern Astrophysical Research (SOAR) telescope, which is a joint project of the Ministério da Ciência, Tecnologia e Inovações do Brasil (MCTI/LNA), the US National Science Foundation’s NOIRLab, the University of North Carolina at Chapel Hill (UNC), and Michigan State University (MSU).

\facilities{NEOWISE, Keck:I (LRIS), Keck:II (NIRES), Magellan:Baade (FIRE), Magellan:Clay (LDSS-3), SOAR (Goodman, Spartan), Hale (WIRC), PO:1.2m (ZTF), ATLAS, PS1, Spitzer}

\bibliography{sample631}{}
\bibliographystyle{aasjournal}

\newpage
\appendix
\restartappendixnumbering

\section{Archival light curve processing}
\label{sec:archival}
While a faint optical transient is reported at the WISE position in the ZTF alert stream as ZTF18abxgjqx\footnote{\url{https://lasair-ztf.lsst.ac.uk/objects/ZTF18abxgjqx/}}, the source is marked as a negative transient, indicating that the transient flux contaminates the ZTF reference images. We therefore derived the ZTF light curve by performing forced difference photometry \citep{Masci2023} followed by addition of the source flux reported in the ZTF reference image catalog. We directly use difference photometry fluxes from the ATLAS forced photometry server\footnote{\url{https://fallingstar-data.com/forcedphot/}} since the reference images were acquired earlier. In both cases, we stack the photometry in time bins of $\approx 7$\,d to improve the signal to noise ratio. We clearly detect a faint optical transient in the archival light curves detected between 2015 and 2023. We also retrieved reduced images from the PanSTARRS survey \citep{Chambers2016} obtained between 2009 and 2014, and performed forced difference imaging photometry using a custom pipeline \citep{De2022}. We only obtain non-detections in PS1 images.

\section{Follow-up imaging and spectroscopy}
\label{sec:followup}
On UT 2022-09-08, we obtained one epoch of $H$ and $Ks$ band NIR imaging of \wtps\ using the Wide-field Infrared Camera (WIRC; \citealt{Wilson2003}) on the Palomar 200-inch telescope. A series of dithered exposures amounting to a total exposure time of $330$\,s were obtained in each filter. The data were reduced and calibrated to the 2MASS system using a custom pipeline \citep{De2020}. While a faint point source is detected at $K = 18.50 \pm 0.20$\,mag, no source is detected in $H$-band to a $5\sigma$ limiting magnitude of $H > 19.6$\,mag. On UT 2023-01-07, $J$-band images were obtained with the Spartan NIR camera \citep{Loh2012} on the 4.1 m Southern Astrophysical Research (SOAR) Telescope (SOAR 2022B-005; PI: De). We obtained a set of dithered images amounting to a total exposure time of $2160$\,s, and reduced the data using a custom pipeline \citep{De2020}. No source is detected at the transient position to a $5\sigma$ limit of $J > 19.0$\,mag. We obtained one epoch of optical imaging on UT 2022-11-28, using the Goodman High Throughput Spectrograph \citep{Clemens2004} on the SOAR telescope. We obtained four exposures amounting to a total exposure time of $1200$\,s in $r$-band. The data were reduced, calibrated and processed through image subtraction using a custom pipeline \citep{De2020} using templates from the Legacy Survey \citep{Dey2019}. At the position of the mid-IR transient, we clearly detect a faint point source with a magnitude of $r = 22.80 \pm 0.15$\,mag.

On UT 2022-10-15, we obtained NIR spectrum of \wtps\ as part of our ongoing NASA Keck program on the NIRES \citep{Wilson2003} instrument (2022B\_N187; PI: De). We obtained a series of dithered exposures amounting to a total exposure time of $\approx 2400$\,s, in addition to a nearby telluric standard star. The data were reduced and calibrated using the \texttt{pypeit} package \citep{Prochaska2020}. On UT\,2023-05-27, we obtained an optical spectrum of the source using the LDSS-3 spectrograph on the Magellan/Clay telescope. The data were obtained using the VPH-Red grism for a total exposure time of $\approx 45$\,minutes. The data were reduced using the \texttt{pypeit} package \citep{Prochaska2020}, and we clearly detect a broad H$\alpha$ emission line at the transient position. On UT\,2023-07-23, we obtained an optical spectrum of the source using the Low Resolution Imaging Spectrometer (LRIS; \citealt{Oke1995}) on the Keck-I telescope (PI: Fremling), for a total exposure time of $\approx 2400$\,s. The data were reduced and extracted using the \texttt{Lpipe} code \citep{Perley2019}. In addition to \wtps, we obtained NIR spectra using the Magellan Baade/FIRE spectrograph \citep{Simcoe2013} of two similar MIR SNe identified in NEOWISE -- SN\,2020jfv and SN\,2020yzq (Section \ref{sec:discussion}). SN\,2020jfv was observed on UT 2022-08-18 using the prism mode for a total exposure time of $\approx 900$\,s, while SN\,2020yzq was observed on UT 2023-08-31 using the echelle mode for a total exposure time of $\approx 3600$\,s. The data were reduced using the \texttt{pypeit} package \citep{Prochaska2020}. The spectra are presented in Figure \ref{fig:spec}.

\section{Delayed MIR brightenings in BTS SNe}
\label{sec:mir_bright}
\begin{figure*}

\includegraphics[width=0.33\textwidth]{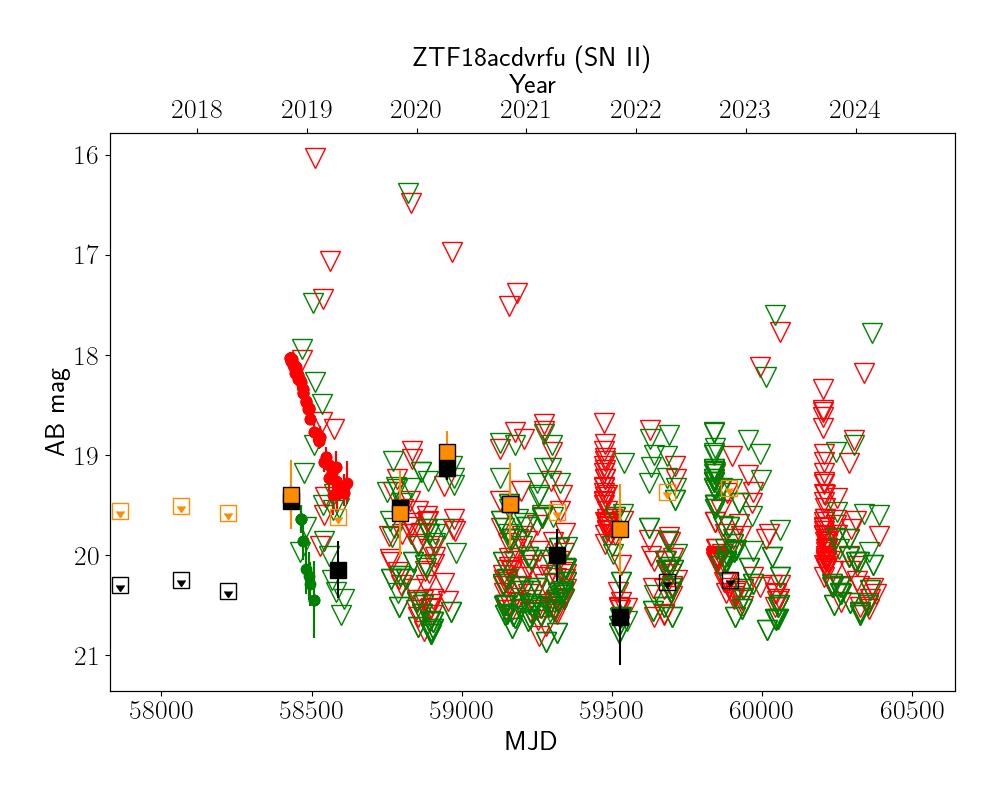}
    \includegraphics[width=0.33\textwidth]{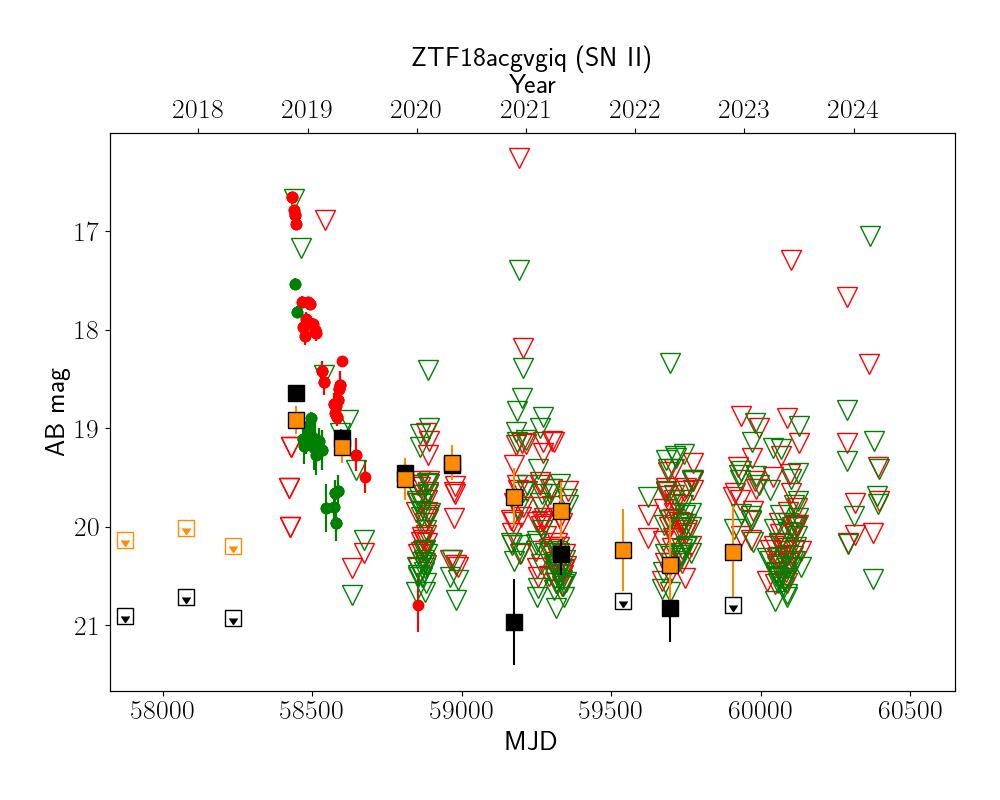}
    \includegraphics[width=0.33\textwidth]{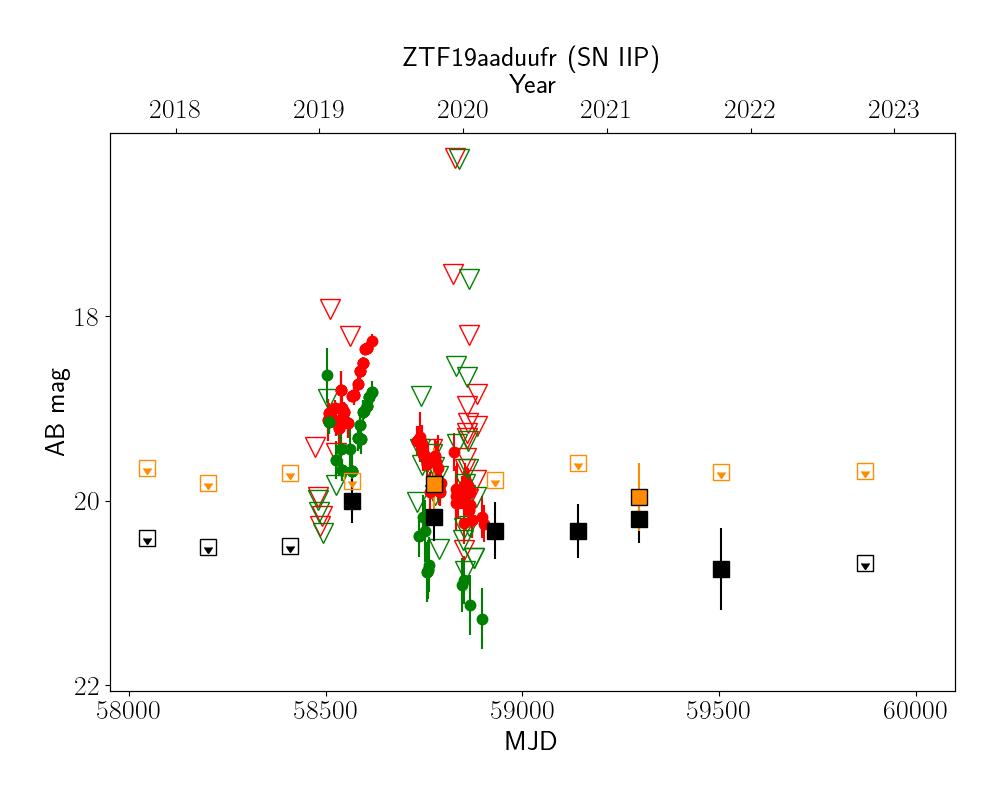}
    \includegraphics[width=0.33\textwidth]{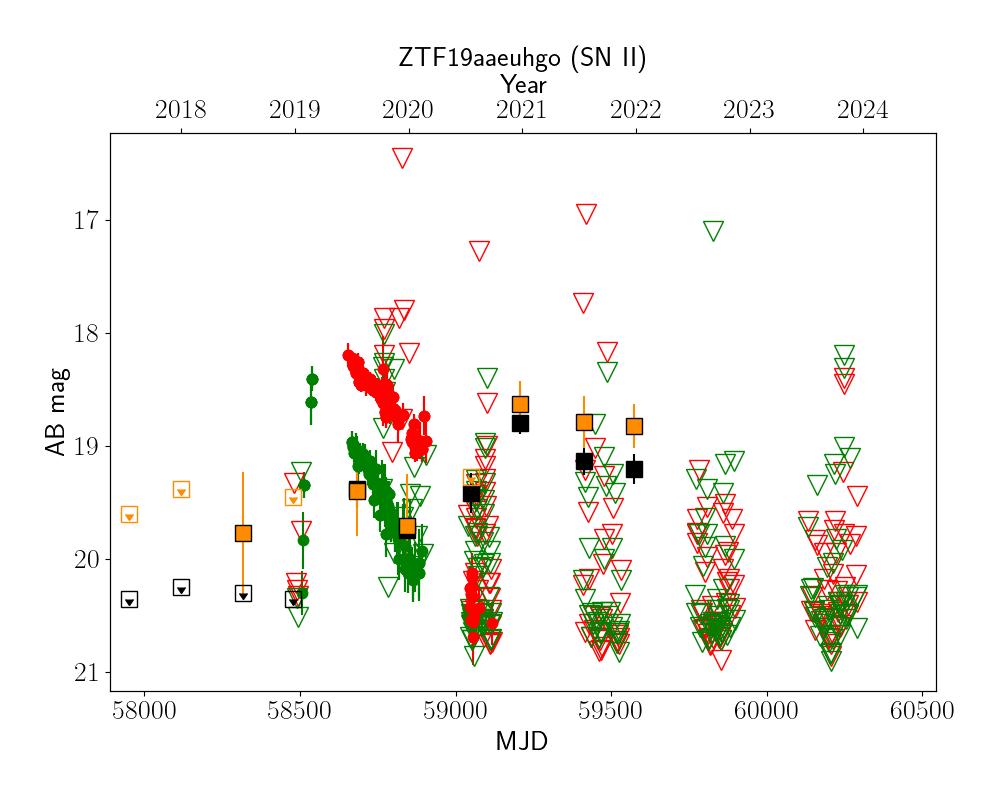}
    \includegraphics[width=0.33\textwidth]{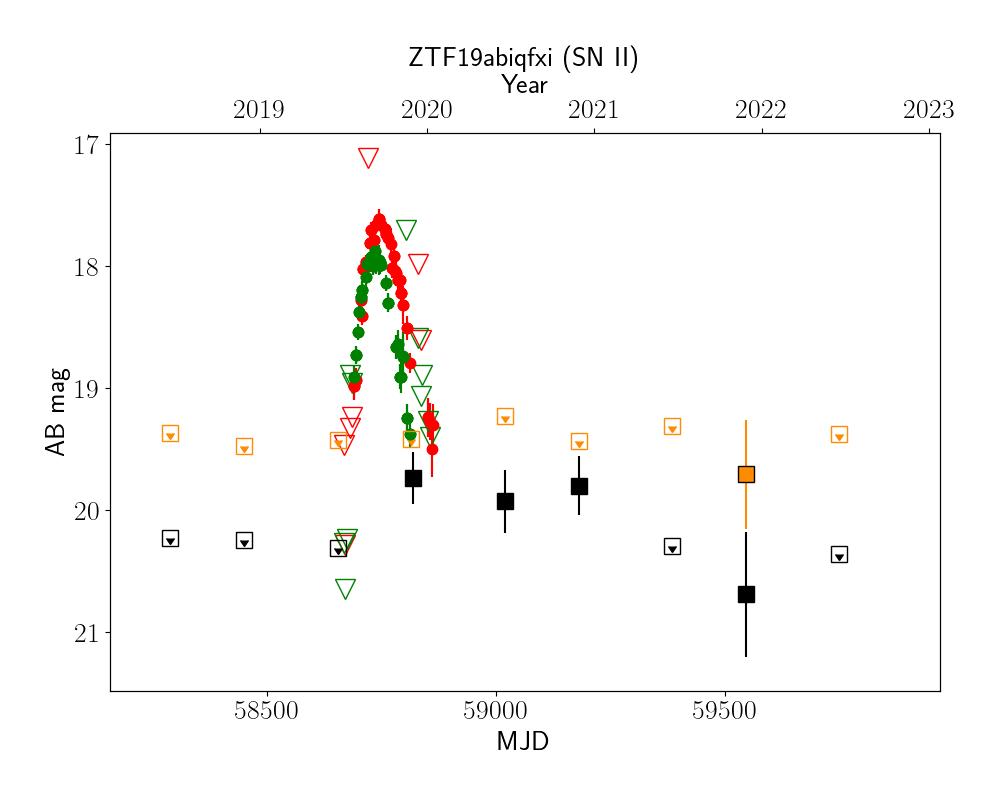}
    \includegraphics[width=0.33\textwidth]{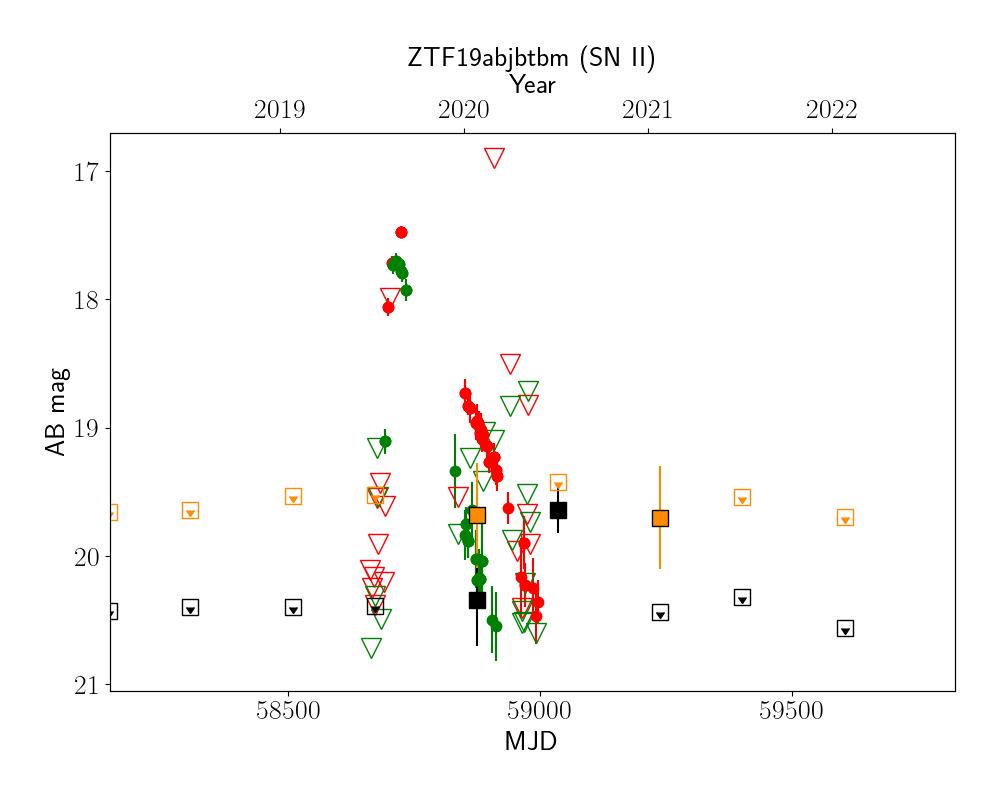}
    \caption{ZTF optical ($g$ and $r$ band shown as green and red symbols respectively) light curves and NEOWISE MIR ($W1$ and $W2$ shown as black and orange respectively) of core-collapse SNe from the Bright Transient Survey that exhibit late-time MIR brightenings in NEOWISE without reported spectroscopic signatures in their nominal classification. We show the nominal peak light spectroscopic classification of the source next to the source name in each panel. Solid symbols denote detections while hollow symbols indicate upper limits.}
    \label{fig:bts_sne_1}
\end{figure*}

\begin{figure*}[!ht]

    \includegraphics[width=0.33\textwidth]{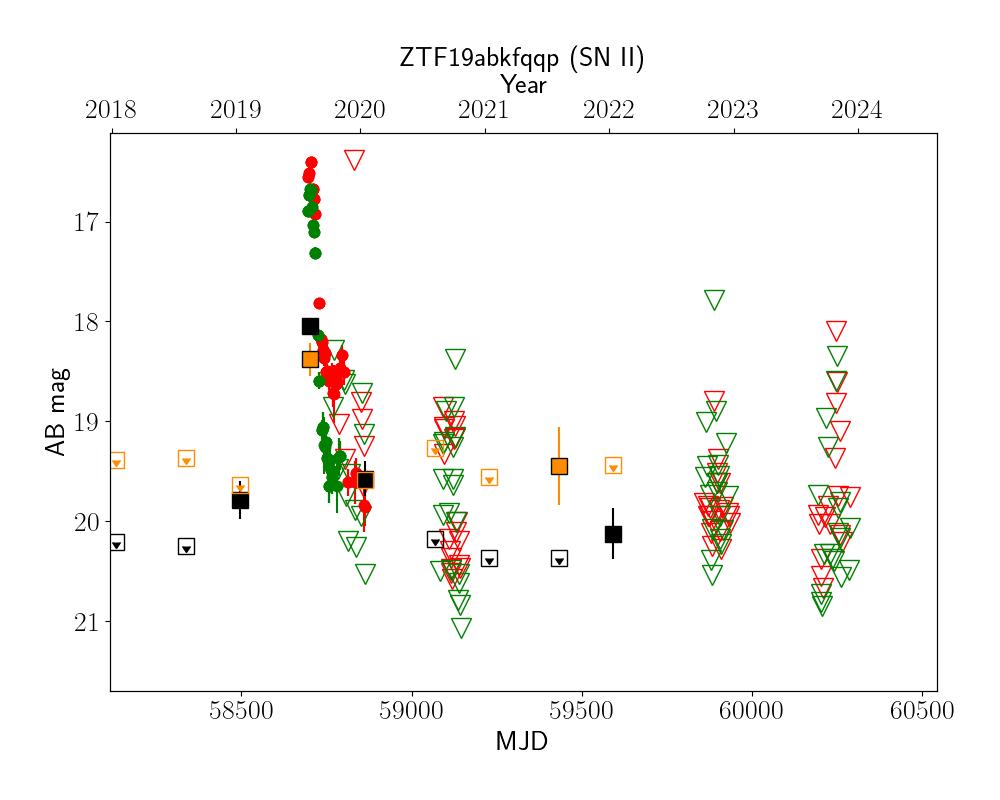}
    \includegraphics[width=0.33\textwidth]{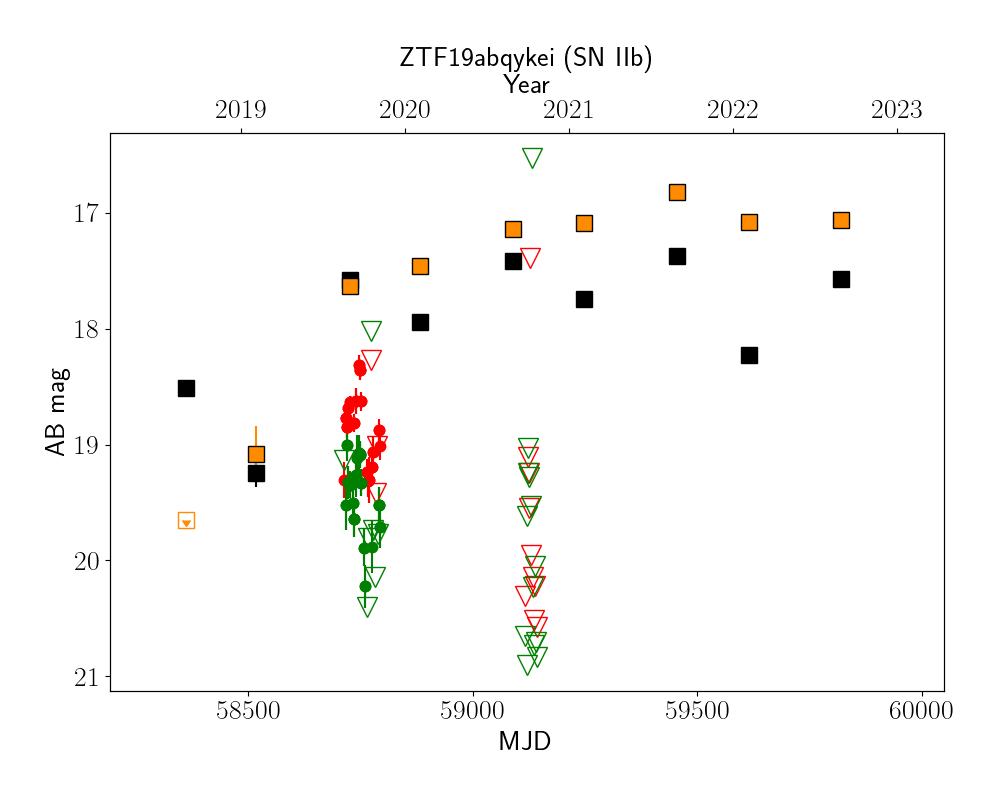}
    \includegraphics[width=0.33\textwidth]{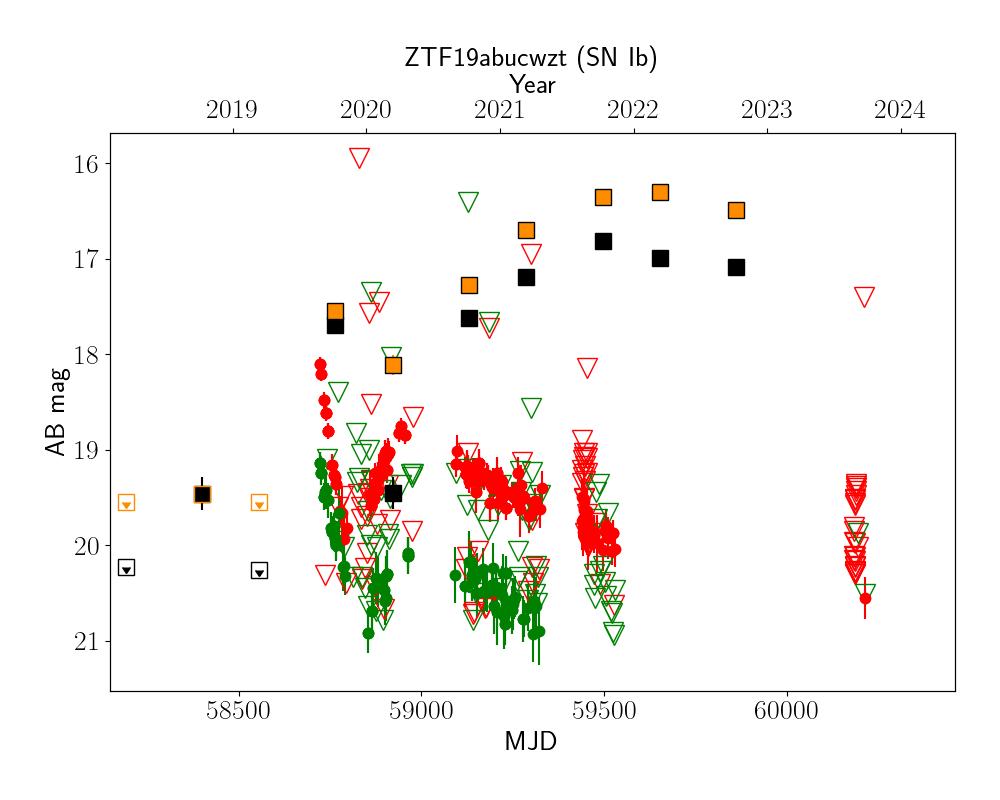}
    \includegraphics[width=0.33\textwidth]{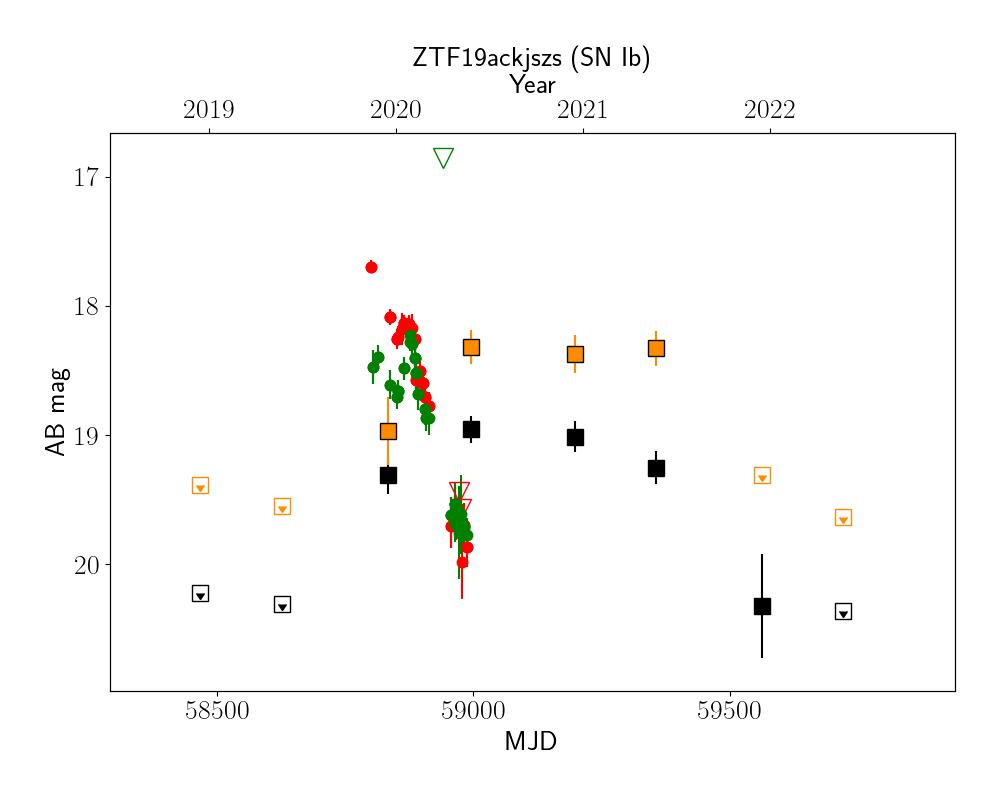}
    \includegraphics[width=0.33\textwidth]{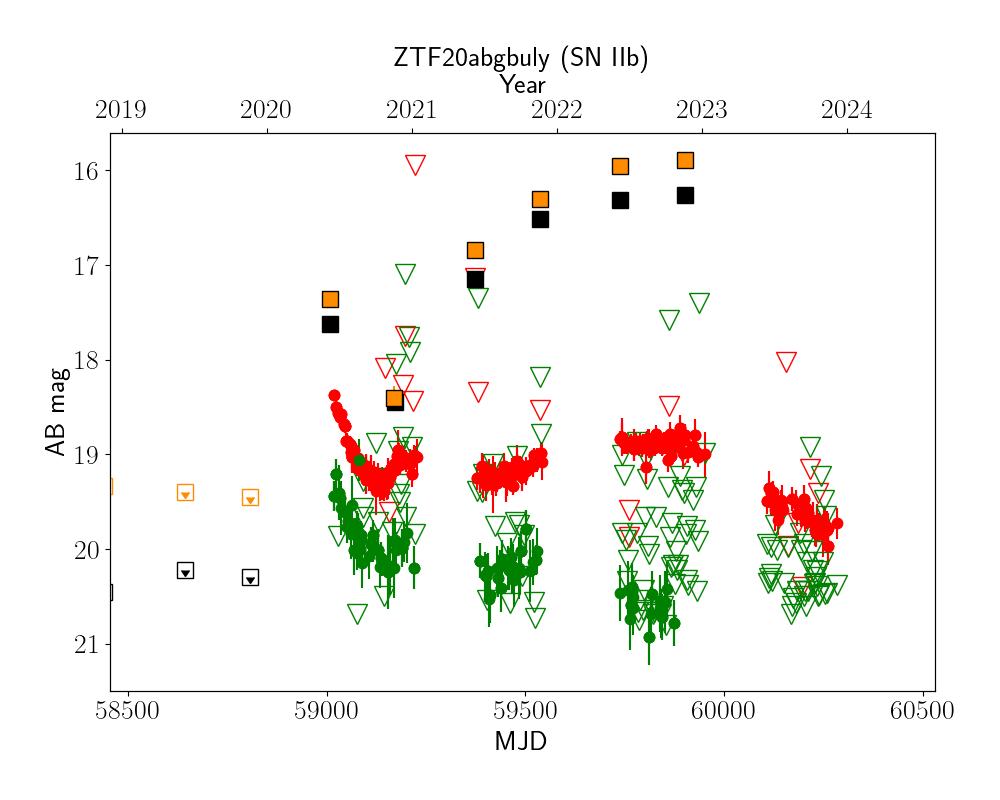}
        \includegraphics[width=0.33\textwidth]{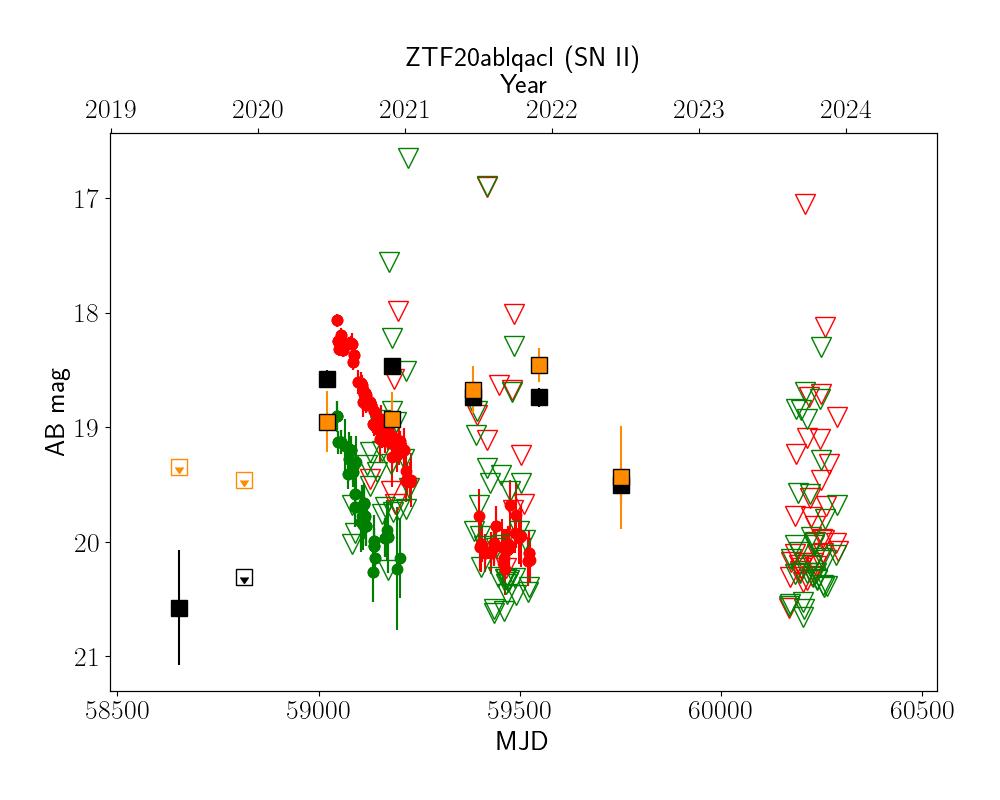}
    \includegraphics[width=0.33\textwidth]{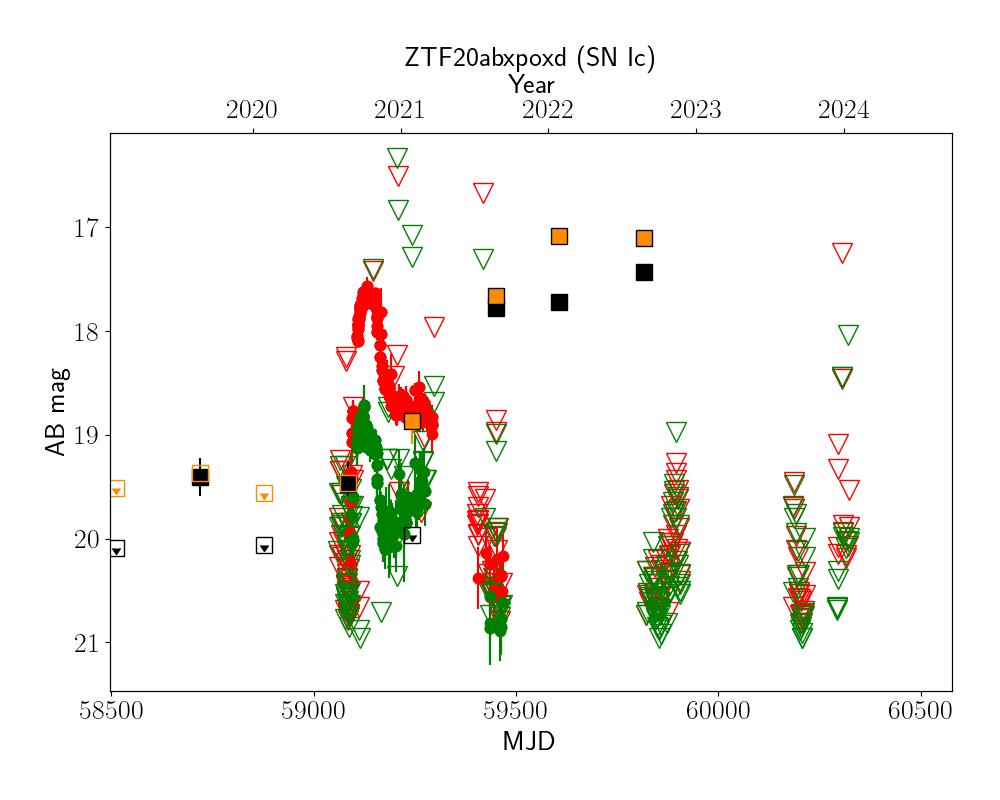}
    \includegraphics[width=0.33\textwidth]{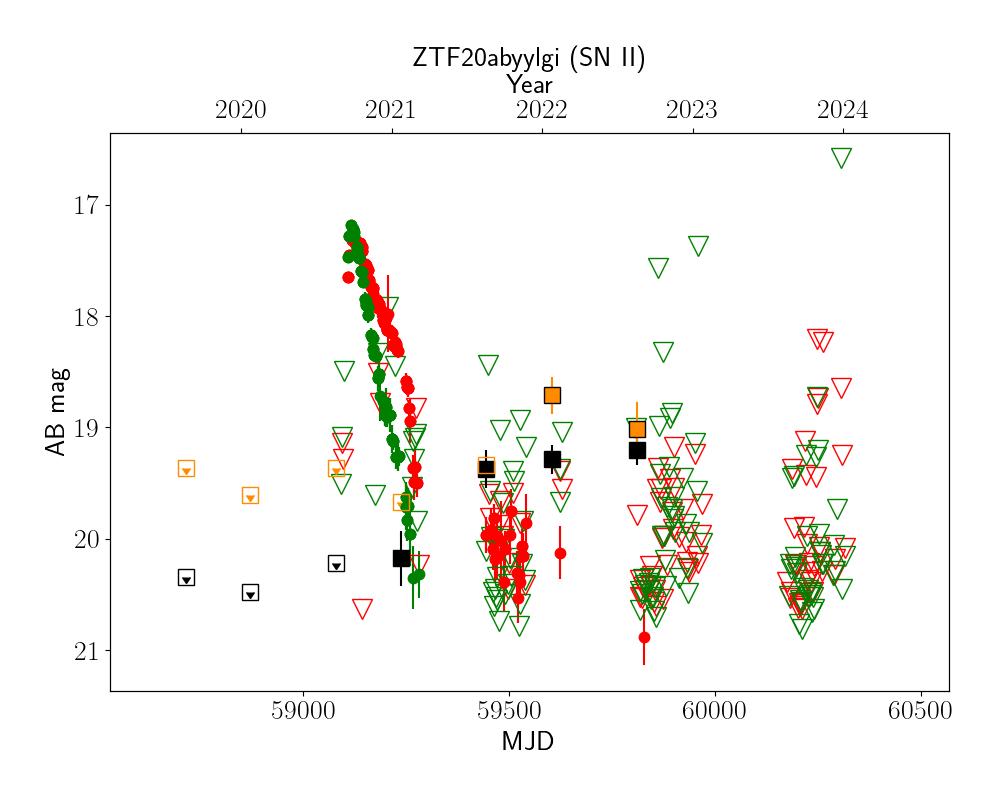}
    \includegraphics[width=0.33\textwidth]{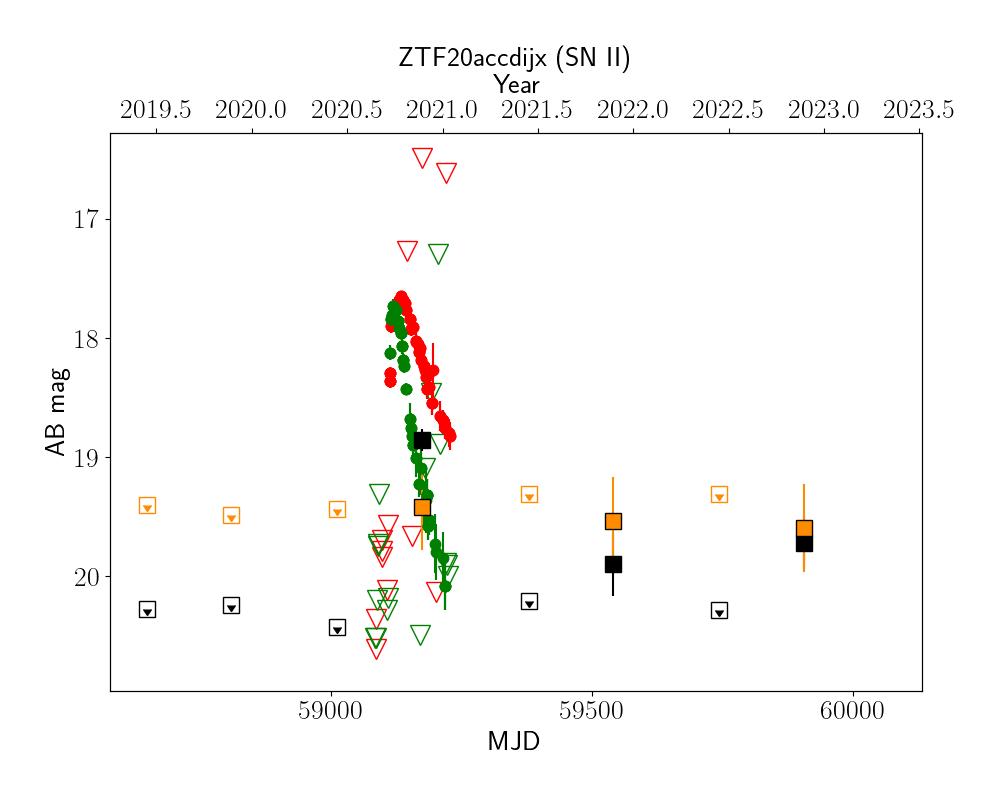}
    \includegraphics[width=0.33\textwidth]{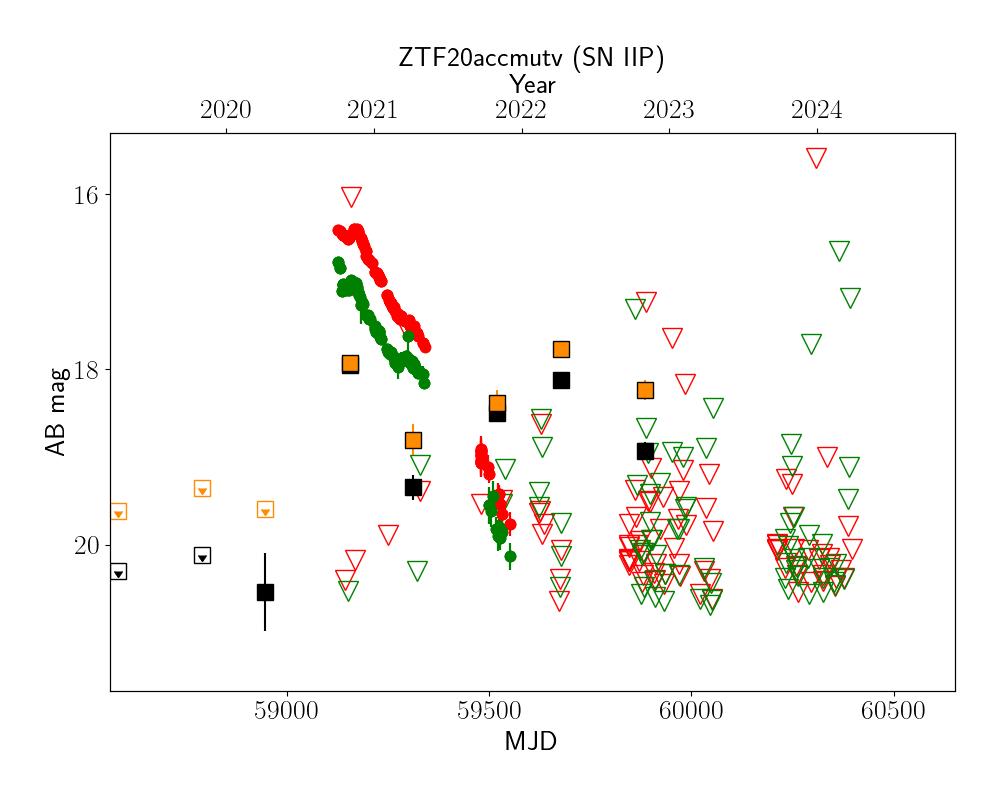}
    \includegraphics[width=0.33\textwidth]{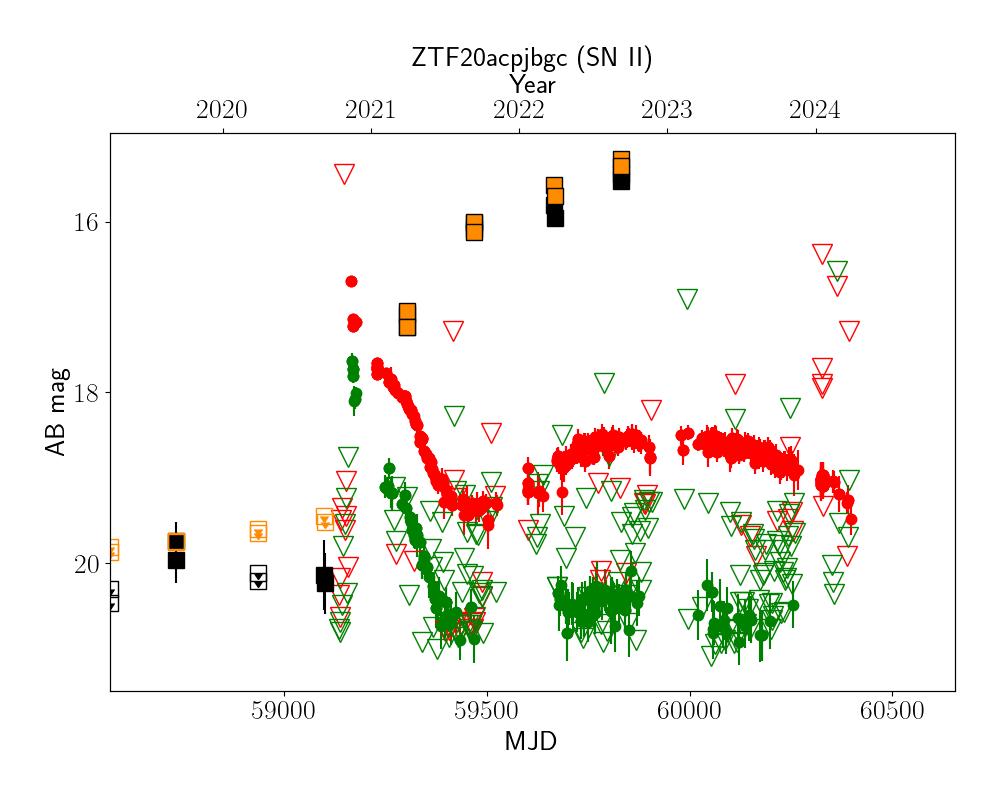}
    \caption{ZTF optical ($g$ and $r$ band shown as green and red symbols respectively) light curves and NEOWISE MIR ($W1$ and $W2$ shown as black and orange respectively) of core-collapse SNe from the Bright Transient Survey that exhibit late-time MIR brightenings in NEOWISE without reported spectroscopic signatures in their nominal classification. We show the nominal peak light spectroscopic classification of the source next to the source name in each panel. Solid symbols denote detections while hollow symbols indicate upper limits.}
    \label{fig:bts_sne_2}
\end{figure*}

In this section, we show optical and MIR light curves of core-collapse SNe from the Bright Transient Survey \citep{Fremling2020, Perley2020} identified to exhibit late-time MIR brightening in NEOWISE. The ZTF photometry was obtained from the publicly available alert packets \citep{Bellm2019} while the NEOWISE photometry is from our forced difference photometry pipeline. The light curves are shown in Figures \ref{fig:bts_sne_1} and \ref{fig:bts_sne_2}. The complete list of objects and their properties are summarized in Table \ref{tab:mir_bright}.

\begin{table}[!h]
    \centering
    \resizebox{\textwidth}{!}{\begin{tabular}{cccccccc}
    \hline
    \hline
         \textbf{ZTF Name} & \textbf{IAU Name} &
         \multicolumn{2}{c}{\textbf{Optical Peak}}&
          \multicolumn{2}{c}{\textbf{NEOWISE Peak}}&
           \textbf{Type} & \textbf{Redshift}  \\
          &  & Mag (band) & $M_{abs}$ & Mag (band) & $M_{abs}$ & &   \\
         \hline
       ZTF\,18acdvrfu  & SN\,2018hov & 18.04 ($r$)& -19.23 & 18.98 (W2) & -18.29 & SN\,II & 0.067  \\
       ZTF\,18acgvgiq  & SN\,2018fru & 16.65 ($r$) & -16.57 & 18.64 (W1) & -14.58 & SN\,II & 0.01026\\
       ZTF\,19aaduufr  & SN\,2019bjr & 18.27 ($r$) & -17.6 & 19.82 (W2) & -16.05 & SN\,IIP & 0.035 \\
       ZTF\,19aaeuhgo  & SN\,2019bcv & 18.20 ($r$) & -19.09 & 18.64 (W2) & -18.65 & SN\,II & 0.0678 \\
       ZTF\,19abiqfxi  & SN\,2019ltz & 17.61 ($r$) & -19.21 & 19.71 (W2) & -17.11 & SN\,II & 0.0542 \\
       ZTF\,19abjbtbm  & SN\,2019lzq & 17.47 ($r$) & -18.69 & 19.64 (W1) & -16.52 & SN\,II & 0.04 \\
       ZTF\,19abkfqqp  & SN\,2019knu & 16.40 ($r$) & -19.14 & 18.04 (W1) & -17.5 & SN\,II & 0.03 \\
       ZTF\,19abqykei  & SN\,2019obh & 18.31 ($r$) & -17.57 & 16.83 (W2) & -19.05 & SN\,IIb & 0.0355 \\
       ZTF\,19abucwzt  & SN\,2019oys & 18.10 ($r$) & -16.15 & 16.31 (W2) & -17.94 & SN\,Ib & 0.01650 \\
       ZTF\,19ackjszs  & SN\,2019tsf & 17.70 ($r$) & -17.84 & 18.33 (W2) & -17.22 & SN\,Ib & 0.03 \\
       ZTF\,20abgbuly  & SN\,2020jfv & 18.38 ($r$)  & -15.95 & 15.89 (W2) & -18.43 & SN\,II & 0.01709 \\
       ZTF\,20ablqacl  & SN\,2020izc & 18.07 ($r$) & -15.61 & 18.46 (W1) & -15.22 & SN\,IIb & 0.01269 \\
       ZTF\,20abxpoxd  & SN\,2020sgf & 17.56 ($r$) & -17.32 & 17.09 (W2) & -17.79 & SN\,Ic & 0.02208 \\
       ZTF\,20abyylgi  & SN\,2020svn & 17.18 ($g$) & -18.28 & 18.72 (W2) & -16.74 & SN\,II & 0.02891 \\
       ZTF\,20accdijx  & SN\,2020twk & 17.65 ($r$) & -17.5 & 18.86 (W1) & -16.29 & SN\,II & 0.025 \\
       ZTF\,20accmutv  & SN\,2020uem & 16.40 ($r$) & -19.82 & 17.78 (W2) & -18.44 & SN\,IIP & 0.041 \\
       ZTF\,20acpjbgc & SN\,2020yzq & 16.69 ($r$) & -17.33 & 15.27 (W2) & -18.74 & SN\,II & 0.01482 \\

         \hline
    \end{tabular}}
    \caption{Summary of ZTF BTS SNe exhibiting late MIR brightening. For each source, we provide the ZTF name, the IAU name, the peak light spectroscopic type, redshift and peak apparent and absolute magnitude in ZTF and NEOWISE bands.}
    \label{tab:mir_bright}
\end{table}

\end{document}